\documentclass[ reprint,amgsmath,amssymb, aps]{revtex4-2}

\usepackage{graphicx}
\graphicspath{{./figs/}}

\usepackage[colorlinks = true,
            linkcolor = blue,
            urlcolor  = blue,
            citecolor = blue,
            anchorcolor = blue]{hyperref}
\usepackage[dvipsnames]{xcolor}
\usepackage{amsmath}
\usepackage{comment}
\usepackage{textcomp}
\usepackage{inputenc}
\usepackage{dcolumn}%
\usepackage{braket} %
\usepackage{soul}
\usepackage{placeins} %

\begin{document}

\date{September 11, 2023} 
\title{Purely anharmonic charge-density wave in the 2D Dirac semimetal SnP}

\author{Martin Gutierrez-Amigo$^{1,2}$}
\email{amigo.martin@ehu.eus}
\author{Fang Yuan$^{3}$}
\author{Davide Campi$^{4}$}
\author{Leslie M. Schoop$^{3}$}
\author{Maia G. Vergniory$^{5,6}$}
\email{maia.vergniory@cpfs.mpg.de}
\author{Ion Errea$^{2,6,7}$}
\email{ion.errea@ehu.eus}
\affiliation{$^1$Department of Physics, University of the Basque Country (UPV/EHU), 48080 Bilbao, Spain}
\affiliation{$^2$Centro de F\'isica de Materiales (CSIC-UPV/EHU), 20018 Donostia/San Sebasti\'an, Spain}
\affiliation{$^3$Department of Chemistry, Princeton University, Princeton, New Jersey 08544, USA}
\affiliation{$^4$Department of Material Science, University of Milano-Bicocca, Via R. Cozzi 55, I-20125 Milano, Italy}
\affiliation{$^5$Max Planck Institute for Chemical Physics of Solids, 01187 Dresden, Germany}
\affiliation{$^6$Donostia International Physics Center (DIPC), 20018 Donostia/San Sebasti\'an, Spain}
\affiliation{$^7$Fisika Aplikatua Saila, University of the Basque Country (UPV/EHU), 20018 Donostia/San Sebasti\'an, Spain}

\begin{abstract}
Charge density waves (CDWs) in two-dimensional (2D) materials have been a major focus of research in condensed matter physics for several decades due to their potential for quantum-based technologies. 
In particular, CDWs can induce a metal-insulator transition by coupling two Dirac fermions, resulting in the emergence of a topological phase. Following this idea, here we explore the behavior of three different CDWs in a new 2D layered material, SnP, using both density functional theory calculations and experimental synthesis to study its stability. The layered structure of its bulk counterpart, Sn$_4$P$_3$, suggests that the structure can be synthesized down to the monolayer by chemical means. However, despite the stability of the bulk, the monolayer shows unstable phonons at $\Gamma$, K, and M points of the Brillouin zone, which lead to three possible charge-density-wave phases. 
All three CDWs lead to metastable insulating phases, with the one driven by the active phonon in the K point being topologically non-trivial under strain.
Strikingly, the ground-state structure is only revealed due to the presence of strong anharmonic effects. 
This, underscores the importance of studying CDWs beyond the conventional harmonic picture, where the system's ground state can be elucidated solely from the harmonic phonon spectra.
\end{abstract}
\maketitle

\section{Introduction}
Charge density waves (CDWs) are a phenomenon of great interest in modern condensed matter physics, characterized by a ground state in which both the charge density and ionic positions exhibit periodic modulation relative to a high-symmetry phase \cite{gruner_1988}.
As a result of their periodic modulation, CDWs can lead to changes in electrical transport properties \cite{gruner_1985,wang_1983,guo_2022}, the thermal conductivity \cite{kwok_1989,kuo_2001}, and the optical absorption \cite{gruner_1988,wang_2022a}. Moreover, CDWs have been proposed to play a role in novel phenomena such as axion insulators \cite{gooth_2019}, where a CDW couples two Weyl fermions in a Weyl semimetal, driving it into an insulating state \cite{wang_2013,wieder_2020,devescovi_2023}. CDWs have also been used to engineer band structures as they can gap out trivial bands, leaving an ideal Dirac semimetal behind \cite{lei_2021}.

A CDW instability is usually predicted from theoretical first-principles calculations by computing the harmonic phonon spectrum and searching for unstable phonon modes.
The active phonons are then tracked into the possible CDW phases. 
Even if the harmonic approximation may hint the order of the CDW and its associated symmetry breaking, it inevitably fails in the description of CDW physics, since, considering that phonons are temperature independent in this approximation, it can never explain when the CDW melts forming the high-symmetry structure at high temperature.
In fact, a CDW transition can be detected experimentally by  measuring the phonons of the high temperature phase and determining which is the phonon mode that collapses with decreasing temperature driving the CDW transition~\cite{weber_2011a,weber_2011,diego_2021}.
Therefore, a first-principles understanding of CDW physics, including the CDW order, requires non-perturbative anharmonic calculations \cite{diego_2021,zhou_2020a,bianco_2020,bianco_2019,zhou_2020,fumega_2023}. 

Tin phosphide compounds are a promising platform for studying CDWs due to their ability to exhibit a wide range of stable and metastable phases \cite{tallapally_2016,li_2020}.
This is largely due to their layered structure, which allows different layers to be ordered and distorted in various ways.
Sn$_4$P$_3$, in particular, has been reported to be a Dirac semimetal \cite{vergniory_2022,vergniory_2019,bradlyn_2017} .
However, it still exists the possibility that the Dirac fermions couple through a CDW, leading to the opening of a gap in the single-particle energy spectrum that may give raise to the emergence of a topological insulator \cite{lei_2021}. 
Given the lack of stability analysis in the literature and the intriguing implications of coupling Dirac fermions by means of a CDW, bulk Sn$_4$P$_3$ and its possible monolayers offer a promising platform to study the emergence of robust topological insulating phases through CDW instabilities.

In this letter, by employing first-principles calculations within density-functional theory (DFT) and the stochastic self-consistent harmonic approximation (SSCHA)\cite{errea_2014,bianco_2017,monacelli_2018,monacelli_2021} to account for anharmonicity, we determine that two distinct monolayers can be derived from bulk Sn$_4$P$_3$, SnP and Sn$_2$P. Of the two monolayers, SnP displays intriguing phonon instabilities at the K and M points, presenting opportunities for the engineering of new materials with specific properties \cite{lei_2021} by coupling and annihilating the band crossings to achieve an insulating phase. Our analysis of these instabilities yielded two distinct insulating phases, one of which could transition into a topological phase under tensile strain. Additionally, a full anharmonic treatment of the phonon spectra revealed a further instability at the $\Gamma$ point. While the K and M phonons are stabilized by anharmonic corrections at higher temperatures, the $\Gamma$ mode remains unstable, indicating the true ground state. These results emphasize the crucial role of anharmonic corrections, both in their quantity and quality,  in CDW systems. We also experimentally synthesize bulk Sn$_4$P$_3$, confirming the crystal structure first described by Olofsson in 1970 \cite{olofsson_1970}. Despite the presence of the two distinct layers is evident in transmission electron microscope (TEM) images, transport measurements on bulk Sn$_4$P$_3$ crystals do not see any signature of CDWs, in agreement with the absence of phonon instabilities in our calculations.

\section{Bulk $\mathbf{Sn_4P_3}$}
\begin{figure}[t]
    \includegraphics[width = \linewidth]{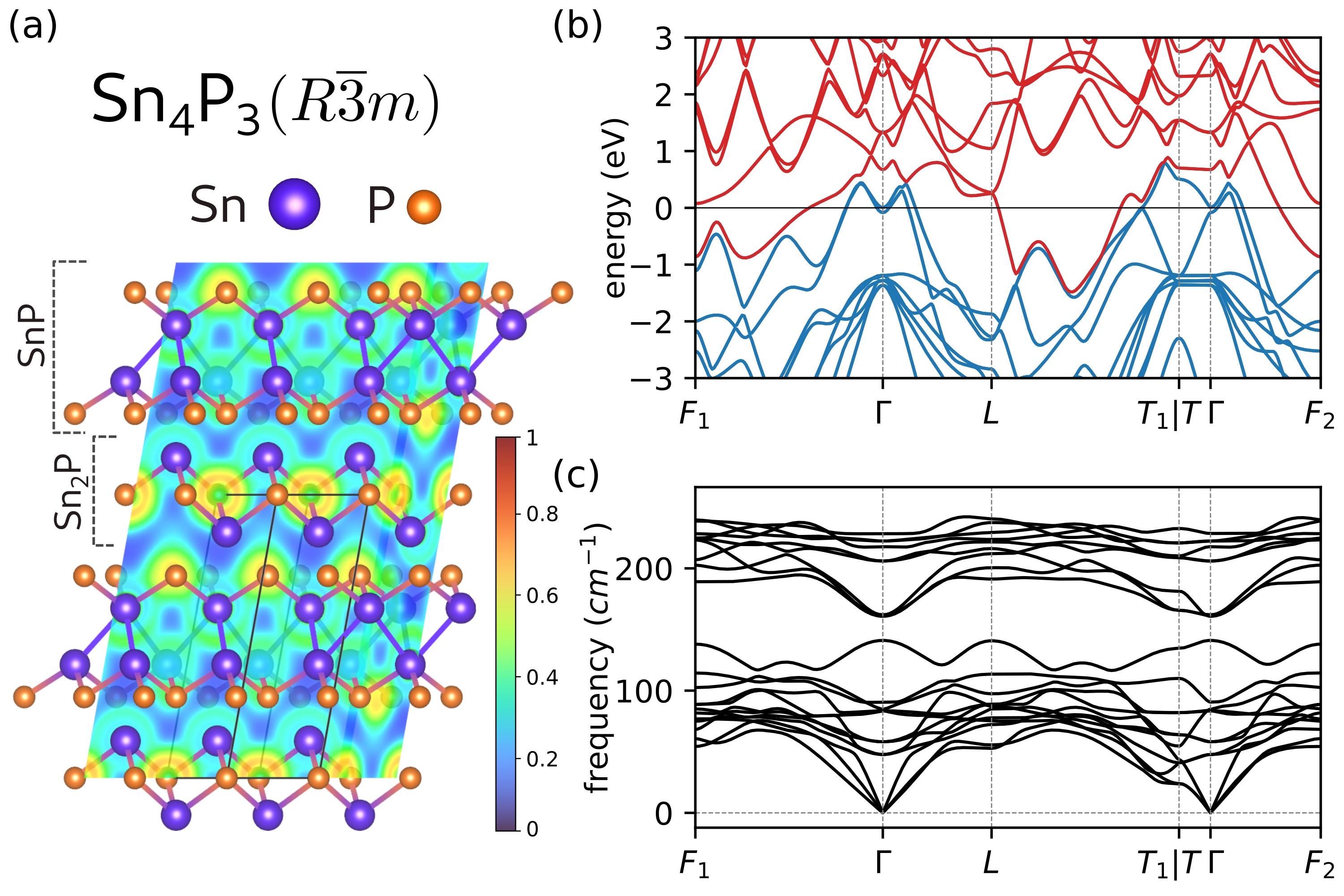}
    \caption{\textbf{Bulk Sn$_4$P$_3$.}
    \textbf{a.} The figure displays the bi-layered structure of Sn$_4$P$_3$, along with an Electron Localization Function (ELF) that reveals the weak Van der Waals coupling between the layers. The two potential monolayers, SnP and Sn$_2$P, have been marked for clarity.
    \textbf{b.} The accompanying electronic band structure displays a unique Dirac crossing at the $L$ high-symmetry point (blue/red colours have been used for valence/conduction bands respectively).
    \textbf{c.} The stability of the system is further confirmed through the harmonic phonon spectra, which shows no indications of a charge-density wave.
    }
    \label{fig_bulk}
\end{figure}

\begin{figure*}
	\includegraphics[width=\linewidth]{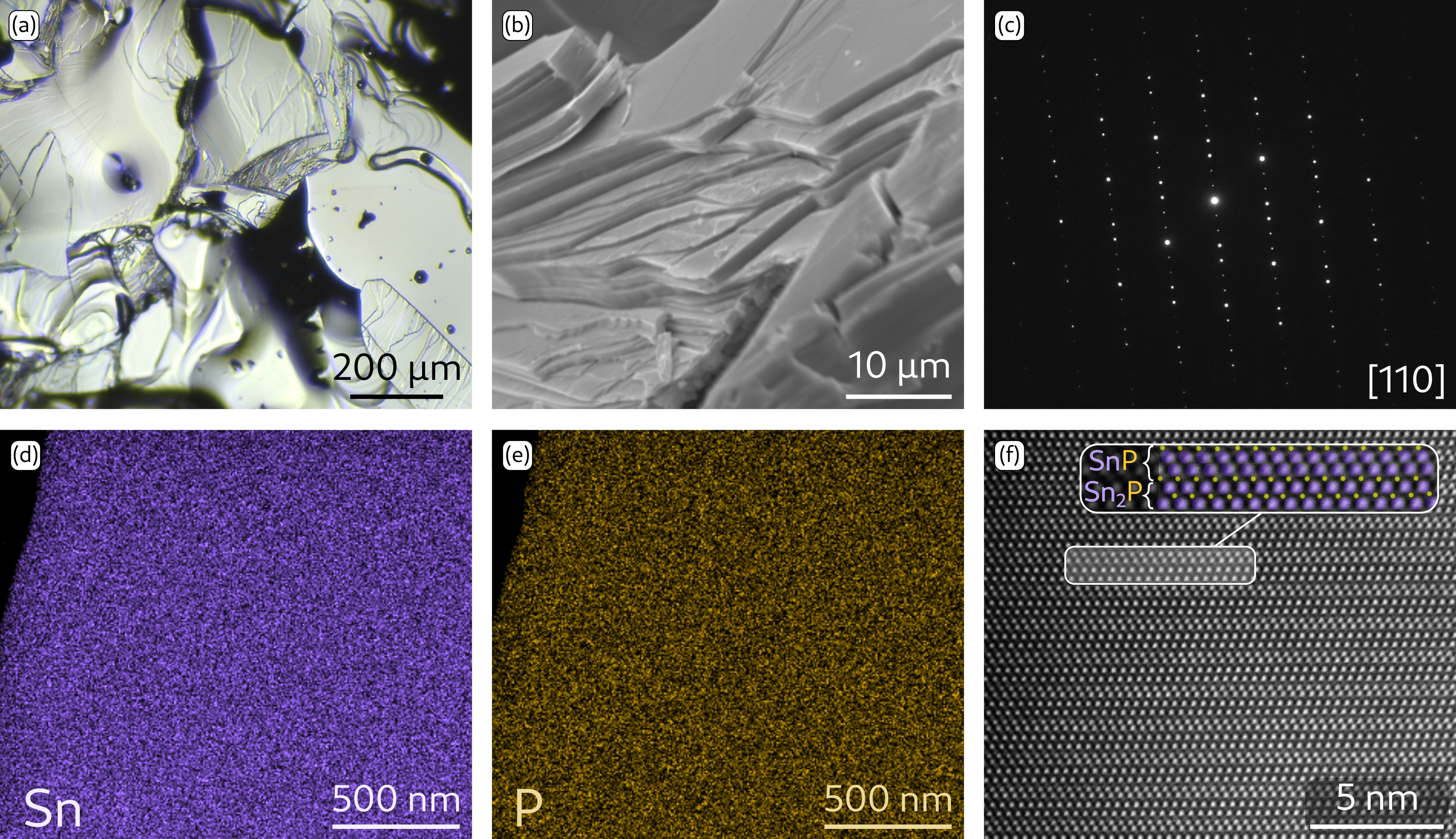}
    \caption{\textbf{ Experimental Result of Sn$_4$P$_3$.} 
    \textbf{a, b.} Optical microscope and scanning electron microscope image of Sn$_4$P$_3$ crystals, showing the layered structure. \textbf{c.} Selected area electron diffraction pattern along [110] axis, suggesting good crystalline quality. \textbf{d, e.} Elemental mapping showing homogeneous Sn and P distribution. \textbf{f.} Atomic-resolution HAADF-STEM image viewed along [110] axis; the inset shows alternating SnP and Sn$_2$P structures.}
    \label{fig_exp}
\end{figure*}

The bulk structure of Sn$_4$P$_3$ (Fig. \ref{fig_bulk}), which has a bilayered structure (space group $R\overline{3}m$), is held together by strong intralayer and  weak interlayer forces.
This weaker interlayer coupling,  together with the fact that the structure holds an odd number of electrons, could favour a layer rearrangement (a CDW) into an insulating phase at sufficiently low temperatures.
To investigate this possibility, we perform first-principles calculations starting from the structure proposed by Olofsson \cite{olofsson_1970}.
First, we relaxed the system to the minimum of the Born-Oppenheimer energy and confirmed the bilayer structure by computing the electron localization function (ELF). As shown in Fig. \ref{fig_bulk}(a), the absence of electronic localization in the interlayer space indicates a weak bonding between the SnP and Sn$_2$P monolayers.
 However, the computed phonon spectrum of the bulk in the harmonic approximation  (see Fig. \ref{fig_bulk}(c)) reveals that the single-band crossing between the valence and conduction bands at the $L$ point (see Fig. \ref{fig_bulk}(b)) remains stable and the bulk does not seem to be prone to a CDW formation.
 In order to test the theoretical analysis outlined, we have successfully synthesized Sn$_4$P$_3$ crystals by heating a mixture of Sn and P in sealed, evacuated quartz tube.
 More experimental details can be found in Appendix \ref{app_experimental}.
 Single-crystal x-ray diffraction indicates the space group $R\overline{3}m$ (No. 166), which is consistent with the previously reported results \cite{olofsson_1970}.
 Full experimentally crystallographic information can be found in Table \ref{tab_SCXRD}.
 Moreover, the obtained structure by single crystal x-ray measurements is in good agreement with the structure used for the theoretical analysis (see Table \ref{tab_Sn4P3}).
 As suggested by the theoretical results, temperature-dependent resistivity measurements confirm that the $R\overline{3}m$ Sn$_4$P$_3$ phase is metallic, showing no signs of a CDW from 1.8 to 400 K (Fig. \ref{RRR}).
 The layered structure can be clearly visualized using both optical microscope (OP) and scanning electron microscope (SEM) (Fig. \ref{fig_exp} (a) and (b)).
 The high-quality crystalline nature of Sn$_4$P$_3$ crystals is confirmed from the selected area electron diffraction (SAED) pattern obtained along the [110] axis using TEM, as illustrated in Fig. \ref{fig_exp} (c)).
 Furthermore, the homogeneous distribution of Sn and P within the Sn$_4$P$_3$ material is confirmed through energy-dispersive spectroscopy (EDS) mapping, as shown in Fig. \ref{fig_exp} (d) and (e).
 Lastly, to confirm the bilayered structure of Sn$_4$P$_3$, atomic-resolution high-angle annular dark-field scanning transmission electron microscopy (HAADF-STEM) imaging was conducted, as depicted in (Fig \ref{fig_exp}(f)).
 The zoom-in inset image reveals that Sn$_4$P$_3$ is composed of alternating SnP and Sn$_2$P layers.

The bilayer structure of bulk Sn$_4$P$_3$ suggests that both SnP and Sn$_2$P monolayers may be synthesized experimentally.
Despite the attempt to exfoliate the different monolayers using  the scotch-tape technique, the exfoliated flakes could not be thinner than $\sim$ 40 nm. (Fig. \ref{fig_AFM}).
This thickness corresponds to about $\sim$ 30 bilayers and is therefore far from the monolayer limit.
As shown in Fig. \ref{fig_exp}(f), the separation between the layers is too small and, hence, might prevent the bulk from delaminating into monolayers. 
This is reflected in the relatively high exfoliation energy from the parent layered material (around 100 meV/$\mathrm{\AA^2}$) computed with vdw-corrected functionals \cite{mounet_2018,campi_2023}. 
However, this does not exclude that these monolayers may be synthesizable by chemical means, as it occurs in for instance in monochalcogenides \cite{chang_2016}.
Another route to explore might be chemical exfoliation, which can access 2D materials that are unobtainable via the scotch tape method \cite{sasaki_1996,yuan_2022,yang_2020}.
We thus further investigate the dynamical stability of the SnP and Sn$_2$P monolayers by first principles.

\section{Monolayers $\mathbf{Sn_2P}$ and $\mathbf{SnP}$}
\subsection{Analysis of the CDW within the harmonic approximation}

The Sn$_2$P monolayer (Fig \ref{fig_monolayer}(a) belongs to space group SG $P\overline{3}m1$ (No. 164) (see Appendix \ref{app_structures} for its structural parameters).
Our calculations reveal that the Sn$_2$P monolayer exhibits a metallic band structure, as illustrated in Fig \ref{fig_monolayer}(b), while its harmonic phonon spectra (see Fig \ref{fig_monolayer}(c)) shows some instabilities near $\Gamma$, corresponding to the out-of-plane acoustic phonon mode.
This result suggests that the Sn$_2$P monolayer may be subject to a rippling instability \cite{fasolino_2007}.
Considering that the band structure of this monolayer is typical of a metal, this rippling instabilities are not expected to yield any significant change to the electronic structure, keeping the structure as metallic.

The band structure of the SnP monolayer is, on the contrary, very interesting.
This monolayer belongs to space group SG $P\overline{3}m1$ (No. 164) and consists of tin and phosphorus atoms at $2d$ and $2c$ Wyckoff positions, respectively.
The structure can be described as a buckled honeycomb of tin atoms sandwiched between two phosphorus triangular lattices, as illustrated in Fig \ref{fig_monolayer}(d).
Its band structure exhibits an interesting single crossing point at the K high-symmetry point, as shown in Fig \ref{fig_monolayer}(e).
Considering that the harmonic phonon spectrum in Fig \ref{fig_monolayer}(f) shows two instabilities at the K and M points, these modes can act as active phonons through which the CDW could couple the Dirac points at K, thereby leading to an insulating state.
To delve deeper into the origin of the CDW, we conducted electron-phonon linewidth and nesting function calculations  (Appendix \ref{app_elec-phonon}), which indicated that the main contribution to the instability arises from the ion-ion interaction.
Consequently, we anticipate a substantial structural reconstruction during the phase transition.

In order to elucidate the low-symmetry structures, we performed a distortion of the SnP monolayer based on the active phonons at the K and M points and relaxed the structure into the minimum of the Born-Oppenheimer energy surface consistent with the symmetry breaking.
The resulting structures belonged to space groups $P\overline{3}1m$ and $P2/c$, respectively, in line with the group theory restrictions for phonons transforming under irreducible representations K$_1$ and M$_1^{-}$~\cite{stokes_1988}, which are the irreducible representations of the active modes. 
Then, we computed the band structure and phonon spectra in both cases, which in turn allows us to analyze both the stability and topology.
Subsequently, we will refer to the resulting structures from the condensation of the K and M modes as SnP-K ($P\overline{3}1m$) and SnP-M ($P2/c$) (see Fig \ref{fig_monolayer} (g  and j)).

\begin{figure*}
    \includegraphics[width = \linewidth]{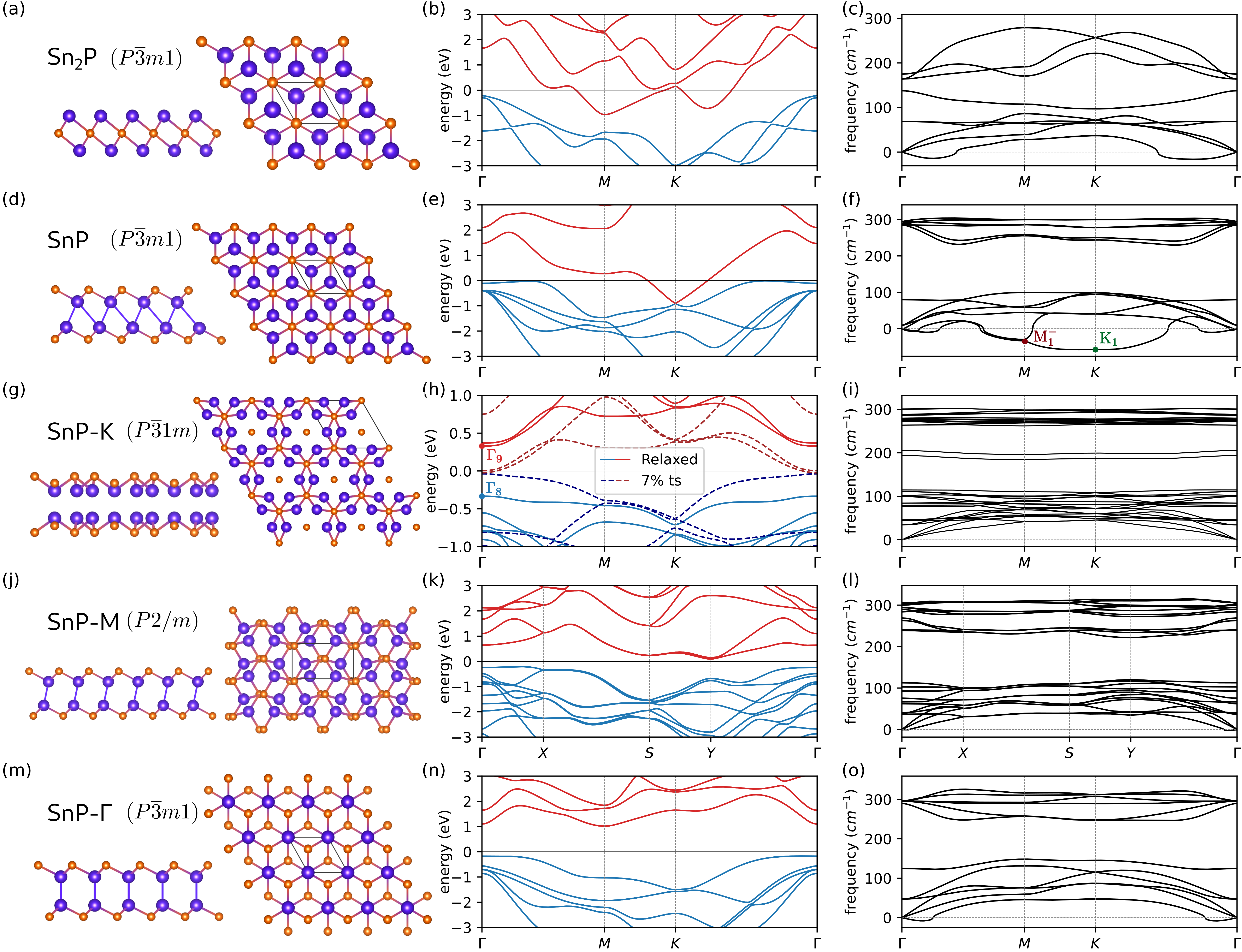}
    \caption{
    \textbf{Sn$_2$P, SnP, SnP-K, SnP-M and SnP-$\Gamma$ monolayers.}
    \textbf{a, d, g, j, m.} Represent the atomic structures of the two original monolayers (Sn$_2$P,SnP) and the CDW phases (SnP-K, SnP-M, SnP-$\Gamma$) from the side and top views.
    \textbf{b, e, h, k, n.} Show the corresponding band structures of each of the monolayers. In \textbf{h}, we demonstrate that by applying an isotropic tensile strain of 7\%  the band gap closes, leading to an inversion of $\Gamma9$ and $\Gamma8$ irreps, and resulting in a topological phase transition.
    \textbf{c, f, i, l, o.} Represent the harmonic phonon spectra of each of the monolayers. As shown in \textbf{f}, the SnP monolayer exhibits two active phonons at the M and K points, respectively. 
    }
    \label{fig_monolayer}
\end{figure*}

Both low-symmetry structures turned out to be insulating phases (see Fig \ref{fig_monolayer}(h,k)), as well as dynamically stable according to the computed harmonic phonon spectra (see Fig \ref{fig_monolayer}(i,l)).
This means that these two structures can be \emph{a priori} stabilized experimentally, showing a large polymorphism of the SnP monolayer.
A topological analysis based on symmetry indicators using topological quantum chemistry (TQC) \cite{bradlyn_2017,vergniory_2019,vergniory_2022} classified both SnP-K and SnP-M as topologically trivial phases.
Nonetheless, it is worth noting that this analysis only verifies  for sufficient conditions for topology and not necessary ones. 
Despite the triviality of the bands of the distorted phases, the insulating gap of SnP-K is remarkably small.
This places this monolayer close to a topological transition:  a band inversion that interchanges the  $\Gamma8$ and $\Gamma9$ irreducible representations at the $\Gamma$ point would lead to a topological phase.
To check this possibility, we perform a more accurate characterization of the gap by carrying out a band structure calculation using the hybrid HSE06 functional~\cite{heyd_2003}.
As a result, we found that under an isotropic tensile strain of 7\% (see Fig \ref{fig_monolayer}(h)), the SnP-K monolayer transitions into a topological insulating phase with the following topological indices $\mathbb{Z}_{2w,3}=1$ and $\mathbb{Z}_4=2$.
Similar strain ranges have previously been realized in different monolayer materials such as graphene \cite{lee_2008} and $\mathrm{MoS_2}$ \cite{bertolazzi_2011}.
Conversely, the SnP-M phase is further away from a topological transition, and strain is not expected to give rise to electronically relevant phases.

\subsection{The impact of anharmonic effects}
\begin{figure}
	\includegraphics[width=\linewidth]{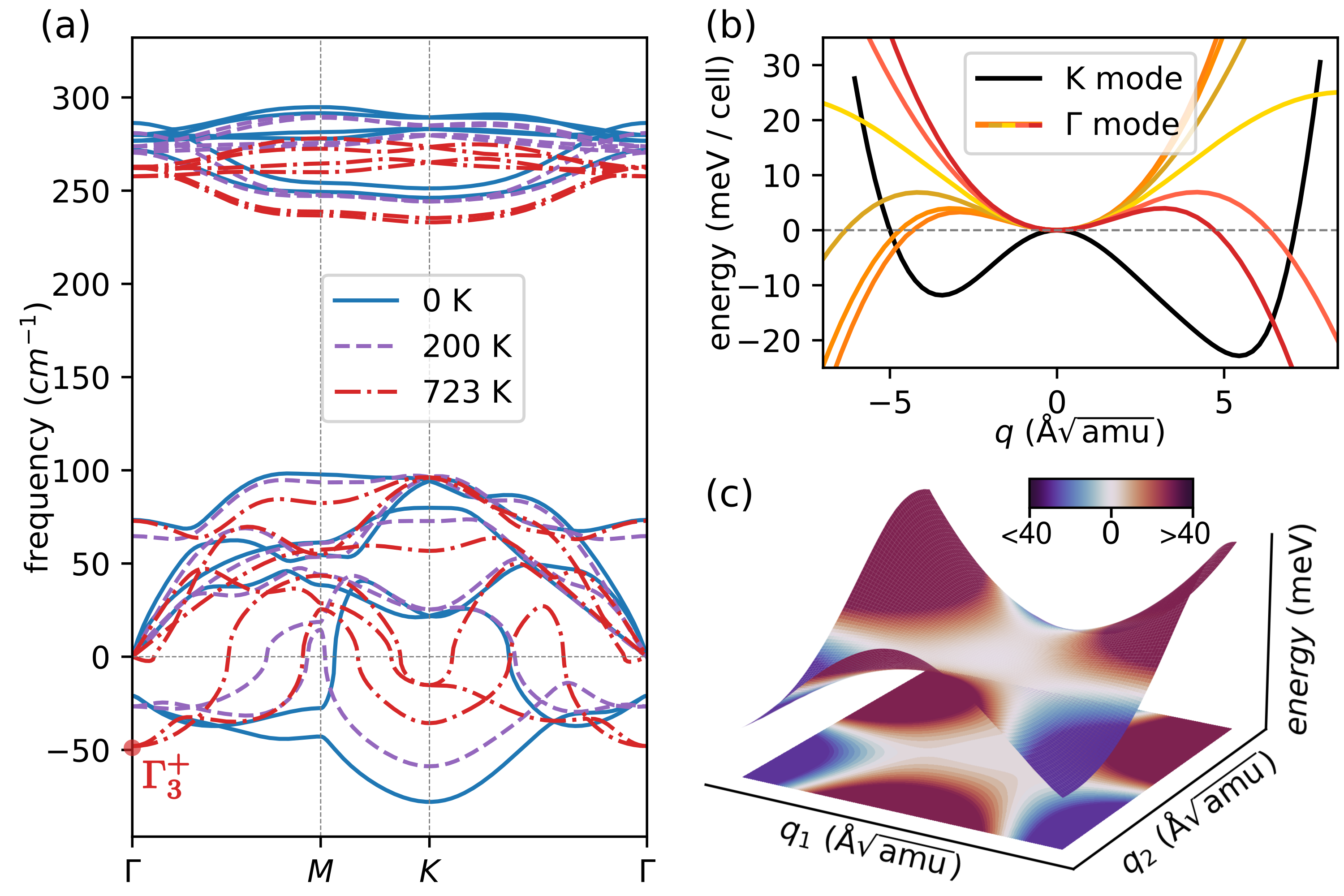}
    \caption{\textbf{Anharmonic phonons of SnP monolayer and Born-Oppenheimer energy surfaces.}
    \textbf{a.} The inset shows the anharmonic phonon behavior at various temperatures. The data reveals that the active phonons located at M and K points exhibit greater stability with rising temperatures, while a new purely anharmonic active phonon emerges at $\Gamma$, which becomes increasingly unstable as the temperature increases.
	\textbf{b.} Born-Oppenheimer energy surface as the system is displaced according to the K and $\Gamma$ active phonons. In the case of the degenerate $\Gamma$ mode, a bidimensional space of deformations is explored through different cuts (see Appendix \ref{app_BO}). The K phonon displays a characteristic "Mexican hat" shape, which becomes more stable as temperature increases, along with quantum fluctuations. Conversely, the $\Gamma$ mode appears as a saddle point with a small potential well.
    \textbf{c.} The complete energy surface of the $\Gamma$ active phonon is shown in this inset. The triple saddle point shape with a small minimum at the center explains why the harmonic approximation fails to capture such instability and why the instability increases with temperature.
    }
    \label{fig_anharmonicity}
\end{figure}

All our analysis so far relies on the harmonic approximation, which is questionable in the vicinity of CDWs. In order to further analyze the stability of the predicted distorted phases, we perform anharmonic first-principles SSCHA calculations of the phonon spectra of both SnP (high-symmetry phase) and SnP-K (low-symmetry phase) at different temperatures. 
As shown in Fig \ref{fig_sscha}, the anharmonic results support the idea of the SnP-K phase being stable after the CDW.
On the other hand, the anharmonic phonons for the SnP monolayer at different temperatures can be seen in Fig \ref{fig_anharmonicity}(a) and present some unexpected outcomes:
both the K and M active phonons stabilize with temperature, with the latter stabilizing within the 0 to 200 K range.
However, although this seems to support the initial idea of the K phonon leading the CDW, the $\Gamma$ point, in fact, shows a new negative degenerate phonon frequency transforming under irreducible representation $\Gamma_3^{+}$. 
This $\Gamma$-point  instability increases with temperature, contrary to the evolution of K and M modes.

To further understand this behavior and the nature of this purely anharmonic active mode, we explore the Born-Oppenheimer energy surface of the SnP monolayer as we distort the structure according to different active modes.
In Fig. \ref{fig_anharmonicity}(b) we show the energy dependence as we displace the ions according to the K active phonon.
The energy surface displays a double-well shape, which explains both the instability of the phonon within the harmonic approximation as well as the anharmonic hardening of the phonon frequency.
Essentially, by increasing the temperature and hence the broadening of the ionic wave function, the potential starts to feel less unstable and more confining.
However, this is not the case for the degenerate phonon that develops an instability at $\Gamma$ within the SSCHA.
As the active phonon is degenerate, this results in a two-dimensional space of deformations that are energetically equivalent just in the harmonic limit.
To analyze such Born-Oppenheimer energy surface, we make use of the three fold symmetry present in the system and compute the energies under different linear combinations of the active modes as described Appendix \ref{app_BO} (Fig \ref{fig_anharmonicity}(b)). 
Then, the whole energy surface can be interpolated as seen in Fig \ref{fig_anharmonicity}(c).
On this occasion, instead of a double-well, the energy surface resembles, rather, a triple saddle point with a small potential well in the middle.
This small well explains why this unstable phonon is invisible in the harmonic approximation.
However, the quantum nature of the ions, even at zero temperature, is enough to escape the potential well, thus leading to a purely anharmonic CDW.
Consequently, the increase of temperature only renders the structure more unstable.

Unlike the active phonons at K and M, the unstable phonon at $\Gamma$ can not be stabilized with temperature, and thus yields to another possible CDW in the SnP monolayer, showing the polymorphism of this 2D material.
We solved the structure arising from the condensation of this degenerate mode by relaxing the lowest energy structure found in the energy surface (Fig \ref{fig_anharmonicity}(c)).
The irreducible representation $\Gamma_3^{+}$ leads to a structure with space group $C2/m$, which after, relaxes back to space group $P\overline{3}m1$.
Despite having the same space group, in the case of SnP-$\Gamma$, tin and phosphorus atoms exchange Wyckoff positions in order to minimize the energy of the system  (see Fig \ref{fig_monolayer}(m)), which again leads to an insulating phase (Fig \ref{fig_monolayer}(n)) classified as trivial according to the TQC analysis.
The harmonic phonons shown in Fig \ref{fig_monolayer}(o) confirm the stability of this SnP-$\Gamma$ phase (which was also confirmed with anharmonic phonons shown in Fig. \ref{fig_sscha}(b)).
Moreover, the energy difference per SnP pair with respect to the high-symmetry phase is the lowest for SnP-$\Gamma$ with $\Delta(\mathrm{SnP-}\Gamma)=-296$ meV, compared to $\Delta(\mathrm{SnP-K/M})=-77/-37$ meV. This indicates that SnP-$\Gamma$ represents the true ground state of the SnP monolayer, while SnP-K and SnP-M are metastable phases.

\section{Conclusions}
In summary, we have studied the stability and properties of Sn$_4$P$_3$ and its monolayers in relation to CDW formations.
The stability of bulk Sn$_4$P$_3$ is confirmed through both density functional theory calculations and experimental synthesis.
The two possible monolayers that this bulk material can yield, SnP and Sn$_2$P, 
show phonon instabilities that may lead to metal-insulator transitions.
The Sn$_2$P monolayer is not expected to change its electronic properties, as it is only subject to rippling instabilities.
However, the SnP monolayer shows a large polymorphism with the formation of different stable and metastable insulating CDW phases, all of which couple the Dirac points of the parent structure.
Remarkably, one of the described phases can be turned into a topological insulator by strain, and the true ground state of the SnP monolayer only is identified thanks to anharmonic effects.
Despite, the predicted polymorphism of SnP could not be identified by exfoliating the bulk, chemical  methods in the few-layers limit may be able to synthesize the anticipated different phases.

\begin{acknowledgments}
We acknowledge fruitful discussions with J. L. Ma\~nes. 
M.G.V., I.E. and M.G.A acknowledge the Spanish Ministerio de Ciencia e Innovacion (grant PID2019-109905GB-C21. 
M.G.A. thanks the Department of Education of the Basque Government for a predoctoral fellowship (Grant no. PRE\_2019\_1\_0304). 
M.G.V. thanks support to the Deutsche Forschungsgemeinschaft (DFG, German Research Foundation) GA 3314/1-1 – FOR 5249 (QUAST) and partial support from European Research Council (ERC) grant agreement no. 101020833. 
This work has also been  funded by the European Union NextGenerationEU/PRTR-C17.I1
, the IKUR Strategy under the collaboration agreement between Ikerbasque Foundation and DIPC on behalf of the Department of Education of the Basque Government
, the Ministry for Digital Transformation and of Civil Service of the Spanish Government through the QUANTUM ENIA project call - Quantum Spain project, 
, as well as by the European Union through the Recovery, Transformation and Resilience Plan - NextGenerationEU within the framework of the Digital Spain 2026 Agenda.
LMS and FY were supported by the  Gordon and Betty Moore Foundation (EPiQS Synthesis Award) through grant GBMF9064. 
We thank Nicola Marzari for making an early version of the MC2D database \cite{campi_2023} and search for CDW available to us.

\end{acknowledgments}

\appendix

\section{\label{app_methods}Methods}
Unless otherwise specified, all first-principles density functional theory (DFT) calculations presented in this study were carried out using {\sc Quantum ESPRESSO} (QE) \cite{giannozzi_2009,giannozzi_2017}, ultrasoft pseudopotentials and a kinetic energy cutoff of 70 Ry with the Perdew-Burke-Ernzerhof parameterization (PBE) approximation for the exchange correlation functional \cite{perdew_1996}.
 The structural relaxations were considered converged when the pressure fell below 0.01 kBar, and a 20x20x1 grid was used. Band calculations were performed with the addition of spin-orbit coupling (SOC) and the same grid.
 Harmonic phonons were computed using density functional perturbation theory (DFPT) within a 6x6x1 supercell in the absence of SOC (in order to avoid unnecessary computational cost).
 The bandstructures computed using hybrid pseudopotentials were done with VASP \cite{kresse_1996,kresse_1996a} in a 9x9x1 grid under the Heyd–Scuseria–Ernzerhof (HSE) approximation with the HSE06 parametrization \cite{heyd_2003}. 
 The anharmonic phonon calculations were done under the Stockastic Self-Consistent Harmonic Approximation (SSCHA) \cite{errea_2014,bianco_2017,monacelli_2018} as implemented in the SSCHA code \cite{monacelli_2021}. 
 In order to capture all the relevant high-symmetry points, the free energy Hessians (SSCHA anharmonic phonons) were done in both 2x2x1 and 3x3x1 supercells in all cases and with the inclusion of the fourth order terms.
 Finally, the topological analysis was done using the IrRep code \cite{iraola_2020} together with tools provided by the Bilbao Crystallographic Server (BCS) \cite{aroyo_2006a,aroyo_2006,aroyo_2011}. 
 
\section{\label{app_experimental}Experimental}
\subsection{Synthesis}
\begin{figure}
\includegraphics[width=\linewidth]{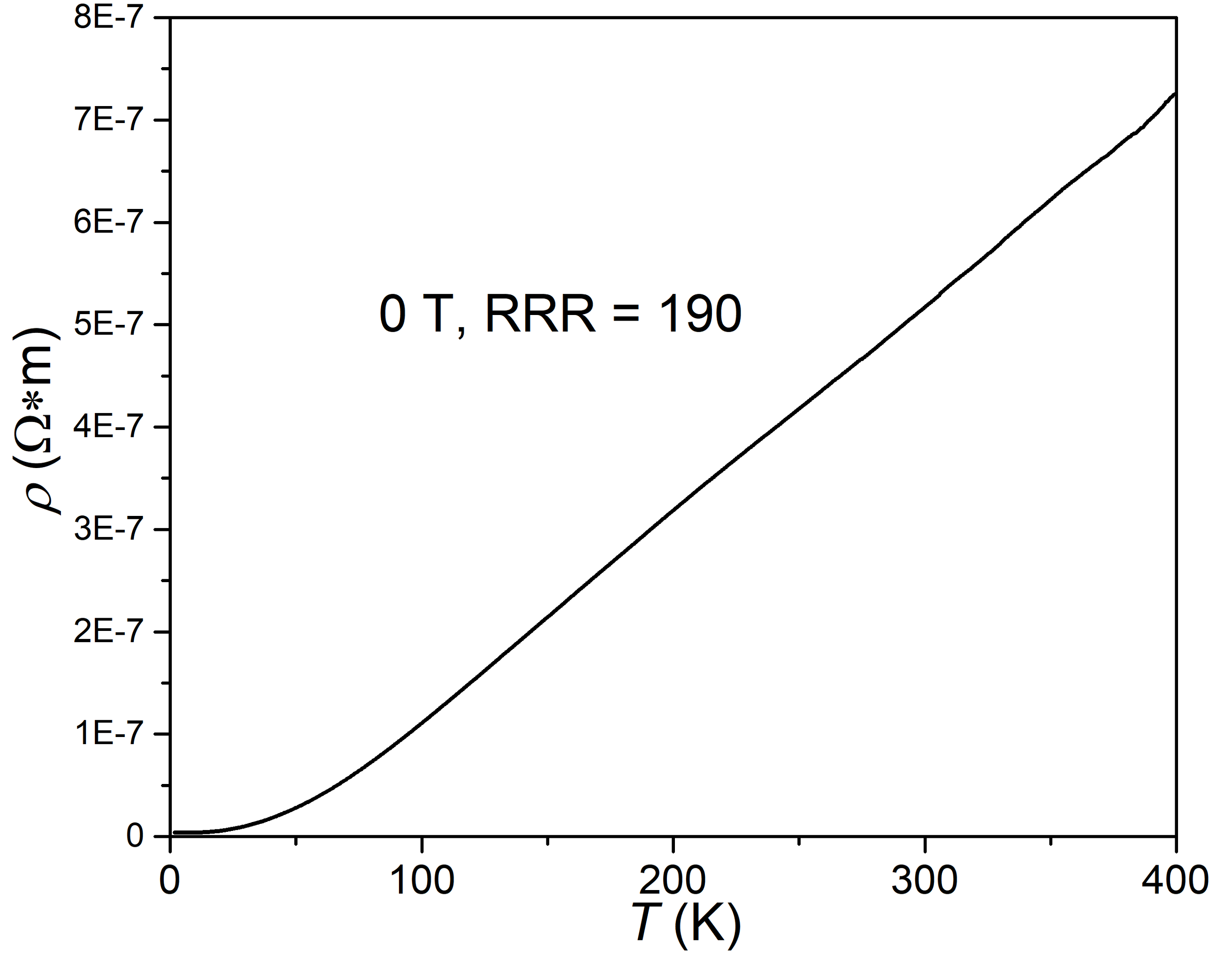}
\caption{\textbf{Temperature-dependent resistivity of Sn$_{4}$P$_{3}$ single crystal, showing a RRR value of 190.}}
\label{RRR}
\end{figure}

The Sn$_{4}$P$_{3}$ bulk crystals were grown by mixing elemental Sn (Sigma-Aldrich 99.99\%) and P (Sigma-Aldrich 99.99\%) in molar ratio and sealed in a evacuated quartz tube, which was then heated to 600 $^{\circ}$C and held at that temperature for 1 week. Subsequently, the mixture was slowly cooled down to room temperature at a rate of 5 $^{\circ}$C/h. This growth method has proven effective in producing high-quality Sn${4}$P${3}$ crystals, as evidenced by a significant residual-resistance ratio (RRR) value of 190 (see Fig. \ref{RRR}). To measure the electrical resistance, the four-probe method was employed using the ac transport option in a Quantum Design physical property measurement system (PPMS). For the electrical contacts, conducting silver paste and gold wire were used, ensuring reliable conductivity during the measurements.

\subsection{Single crystal X-ray diffraction measurements} 
Single-crystal x-ray diffraction (SCXRD) measurements were conducted utilizing a Bruker D8 VENTURE diffractometer equipped with a PHOTON III CPAD detector and a graphite-monochromatized Mo-$K_{\alpha}$ radiation source. The acquired raw data underwent several corrections, including background correction, polarization correction, Lorentz factor correction, and multiscan absorption correction. To solve and refine the crystal structure, the OLEX2 software package was employed, which provides a comprehensive suite of tools for crystallographic analysis and refinement of Sn$_{4}$P$_{3}$. (see Table \ref{tab_SCXRD})

\begin{table}
\begin{ruledtabular}
\renewcommand{\tablename}{Table}
\begin{tabular}{@{}lllll@{}}
Refined  Composition                                           & Sn$_{4}$P$_{3}$                           & &  &  \\ 
Crystal Dimension (mm)                                          & 0.041$\times$0.060$\times$0.078      
&  &  &  \\
Temperature (K)                                                 & 295                                   
&  &  &  \\
Radiation Source, $\lambda$ (\AA)                               &  Mo K$\alpha$, $\lambda$ = 0.71073 \AA  
&  &  &  \\
Absorption Correction                                           & multi-scan                           
&  &  &  \\
Space Group                                                     & $R\overline{3}m$                                  &  &  &  \\
$a$ (\AA)                                                       & 3.971(2)                            
&  &  &  \\
$b$ (\AA)                                                       & 3.971(2)                            
&  &  &  \\
$c$ (\AA)                                                       & 35.397(7)                            
&  &  &  \\
Cell Volume (\AA$^{3}$)                                         & 483.5(5)                         
&  &  &  \\
Absorption Coefficient (mm$^{-1}$)                              & 15.929                                
&  &  &  \\
2$\theta_{min}$ , $\theta_{max}$                                 & 6.906, 135.338                           &  &  &  \\
Refinement Method                                               & Goodness-of-fit on F$^{2}$                               &  &  &  \\
Number of Reflections                                           & 22369                                
&  &  &  \\
Number of Parameters                                            & 13                                    
&  &  &  \\
Independent Reflections                                         & 1215                             
&  &  &  \\
R(I\textgreater{}$2\sigma$), R$_{w}$(I\textgreater{}$2\sigma$)  & 1.35, 3.03                            &  &  &  \\
R(all), R$_{w}$(all)                                            & 1.56, 3.11                            &  &  &  \\
$\Delta\rho_{max}$ , $\Delta\rho_{min}$ (e \AA$^{-3}$)          & 1.16, -1.36                           &  &  & \\ 
\end{tabular}
\caption{\label{tab_SCXRD}Crystallographic Information for Sn$_{4}$P$_{3}$.}
\end{ruledtabular}
\end{table}

\subsection{TEM Sample Preparation}
The lamella sample for cross-sectional transmission electron microscopy (TEM) studies was prepared by cutting it from a bulk single crystal and subsequently thinning it to $\sim$50 nm using Helios DualBeam focused ion beam (FIB)/scanning electron microscope (SEM) system. To minimize surface damage caused by the high-energy FIB, the TEM lamellar samples were polished using a 2 kV gallium ion beam. Once the TEM lamella was obtained, it was carefully transferred to a Mo TEM grid and immediately loaded to the high vacuum chamber to avoid oxygen and moisture attacking under ambient condition.

\subsection{Scanning/Transmission Electron microscopy}
The selected area electron diffraction (SAED) and atomic-resolution scanning transmission electron microscopy (STEM) images were obtained using a Titan Cubed Themis 300 double Cs-corrected scanning/transmission electron microscopy (S/TEM) under a 300 kV operating voltage. This microscope offers a spacial resolution of 0.07 nm and energy resolution of 0.8 eV. Furthermore, it is equipped with a super-X Energy dispersive spectrometry (EDS) system, enabling precise elemental mapping capabilities.

\begin{figure}
	\includegraphics[width=\linewidth]{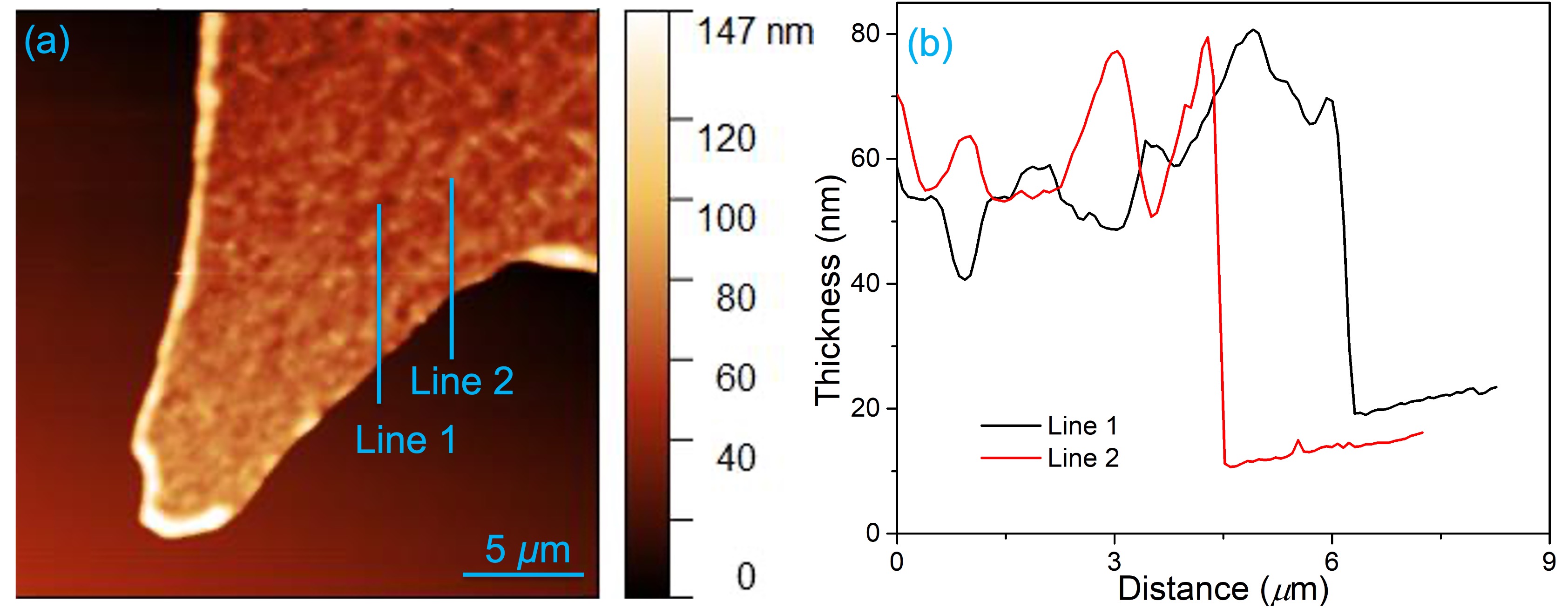}
    \caption{\textbf{Thickness determination of Sn$_4$P$_3$ thin flake obtained using Scotch tape.} \textbf{a} AFM image of a typical Sn$_4$P$_3$ thin flake. \textbf{b.} Height profiles of selected lines on image (a).}
    \label{fig_AFM}
\end{figure}

\subsection{Thickness Determination}
First, the thin flakes of Sn$_{4}$P$_{3}$ were obtained through the process of mechanical exfoliation from crystals, using Scotch tape. The flakes were then transferred onto Si/SiO$_{2}$ wafers. Si/SiO$_2$ wafers were bought from SI-TECH, the thickness of Si and SiO$_2$ is 500-550 $\mu$m and 285 nm, respectively. The thickness of nanosheets was checked on an optical microscope (OM). Upon observing promising nanosheets under the OM, the nanosheet thickness was measured with a Bruker Dimension ICON3 Atomic Force Microscope (AFM) operating in a tapping mode. The acquired raw data were processed using Gwyddion software package. (Fig. \ref{fig_AFM})

\section{\label{app_BO}Calculation of the Born-Oppenheimer energy surface.} 
In order to compute the Born-Oppenheimer (BO) energy surface, we calculate the energy by displacing the ions according to the corresponding active phonons. 
In the harmonic approximation, the displacement $\vec{u}$ of atom $i$ in the unit cell $\vec{R}$ can be expressed as:
$$\vec{u}_i(\vec{R})=\mathrm{Re}\{ \sum_{s \vec{k}} q_s (\vec{k}) \frac{\vec{\varepsilon}_i^{s}(\vec{k})}{\sqrt{M_i}}e^{i\vec{k}\vec{R}} \}$$
Here, $s$ labels the mode, $M_i$ represents the ionic mass of atom $i$, $\vec{\varepsilon}_i^s(\vec{k})$ is the polarization vector, and $q_s(\vec{k})$ is the order parameter associated with the $s$ mode at wave number $\vec{k}$. 
By plotting the energy against the order parameter $q$, we obtain the BO energy surface along that specific direction in the order parameter space.

For the degenerate phonon at $\Gamma$, which possesses a bidimensional order parameter space $(q_1, q_2)$, we perform calculations in six different directions within this space. 
To sample this region thoroughly, we evaluate the order parameter along the directions $(\cos{(\alpha)}q_1, \sin{(\alpha)}q_2)$, where $\alpha$ takes values of $0$, $\frac{\pi}{18}$, $\frac{\pi}{9}$, $\frac{\pi}{6}$, $\frac{2\pi}{9}$, and $\frac{5\pi}{18}$. 
This sampling strategy effectively covers 120 degrees of the order parameter space. 
Then, by taking advantage of the system's three-fold symmetry, we can obtain a complete sampling of the $(q_1, q_2)$ space.

\section{\label{app_SSCHA}SSCHA calculations of CDW structures.} 
\begin{figure}
	\includegraphics[width=\linewidth]{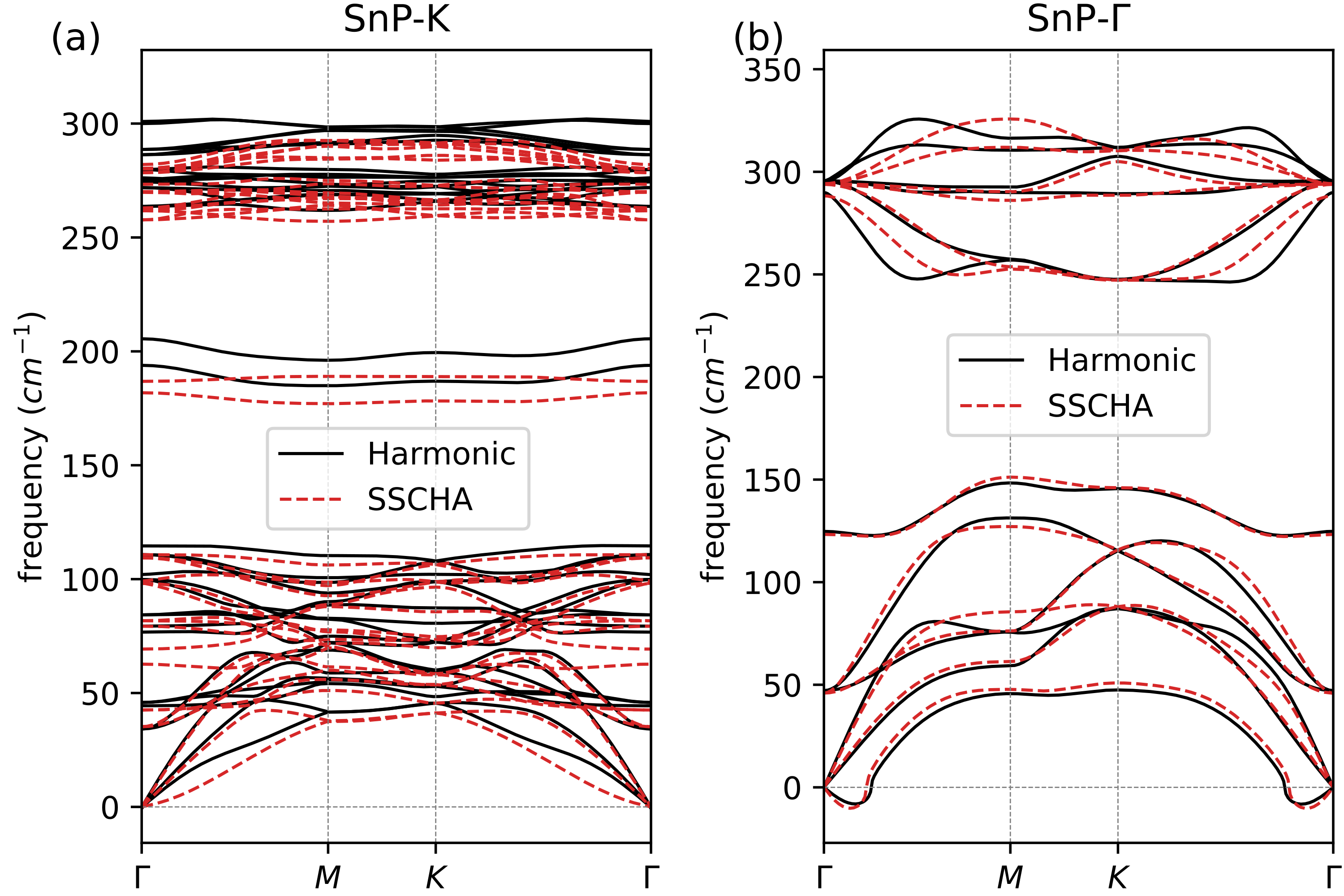}
    \caption{
    \textbf{Anharmonic phonon spectra of SnP-K and SnP-$\Gamma$ monolayers.}
    \textbf{a} and \textbf{b} show the anharmonic and harmonic phonon spectrums of SnP-K and SnP-$\Gamma$ respectively. In both cases the SSCHA calculation has been done at T = 0 and the plotted spectra corresponds to the 3x3x1 grid.
    }
    \label{fig_sscha}
\end{figure}

The stability of the SnP-K and SnP-$\Gamma$ charge-density wave structures was confirmed also by computing the anharmonic phonon dispersions in both a 2x2x1 and 3x3x1 grids (see Fig. \ref{fig_sscha}).

\section{\label{app_elec-phonon}Electron-phonon linewidth and nesting function}
The electron-phonon matrix elements $g_{n\mathbf{k},m\mathbf{k+q}}^{\mu}$ for a phonon mode $\mu$ with momentum $\mathbf{q}$ and two electronic states in bands $n$ and $m$ with electronic momenta $\mathbf{k}$ and $\mathbf{k+q}$ are calculated withing DFPT as:
\[
g_{n\mathbf{k},m\mathbf{k}+\mathbf{q}}^{\mu}=\sum_{s\alpha}
\frac{1}{\sqrt{ 2M_{s}w_{\mu}(\mathbf{q}) }}\varepsilon _{\mu s}^{\alpha} (\mathbf{q})
\braket{ n\mathbf{k} | \left[ \frac{ \partial V_{KS} }{ \partial u_{s}^{\alpha}(\mathbf{q}) } \right]_{0}  |m\mathbf{k}+\mathbf{q} } 
.\] 
where $M_{s}$ is the atomic mass of atom $s$, $w_{\mu}(\mathbf{q})$ is the frequency of the mode, $\varepsilon_{\mu s}^{\alpha}(\mathbf{q})$ is the polarization vector, with $\alpha$ being a Cartesian direction and $\braket{ n\mathbf{k} | \left[ \frac{ \partial V_{KS} }{ \partial u_{s}^{\alpha}(\mathbf{q}) } \right]_{0}  |m\mathbf{k}+\mathbf{q} } $ are the matrix elements of the derivative of the Kohn-Sham potential with respect to the atomic displacements of the phonon mode.
Then, the electron-phonon contribution to the phonon linewidth for mode $\mu$ with momentum  $\mathbf{q}$ can be calculated as:
\[
HWHM_{elph,\mu}(\mathbf{q})=\frac{2 \pi w_{\mu}(\mathbf{q})}{N_{k}}\sum_{\mathbf{k}n m}
\left| g_{n\mathbf{k},m\mathbf{k+q}}^{\mu} \right| \delta(\epsilon _{n\mathbf{k}})\delta(\epsilon _{m\mathbf{k+q}})
.\] 
with $N_{k}$ being the number of $\mathbf{k}$ points in the sum and $\epsilon_{n\mathbf{k}}$ the energy of the state $\ket{n\mathbf{k}}$ measured from the Fermi level. Notice that $HWHM_{elph,\mu}$ is independent from the frequency $w_{\mu}(\mathbf{q})$, since this term cancels with the one from $g_{n\mathbf{k},m\mathbf{k+q}}^{\mu}$, which allows us to define it even for negative frequencies.
The $HWHM_{elph,\mu}$ was computed using a $48 \times\ 48 \times\ 1$ grid and a Gaussian smearing of 0.003 Ry for the Dirac deltas.
The nesting function $\zeta(\mathbf{q})$ was computed using the same grid and Gaussian smearing and defined as:
\[
\zeta\left( \mathbf{q}\right)=\frac{1}{N_{k}} \sum_{\mathbf{k}nm}\delta(\epsilon _{n\mathbf{k}})\delta(\epsilon _{m\mathbf{k+q}})
.\] 

\begin{figure}
\includegraphics[width=\linewidth]{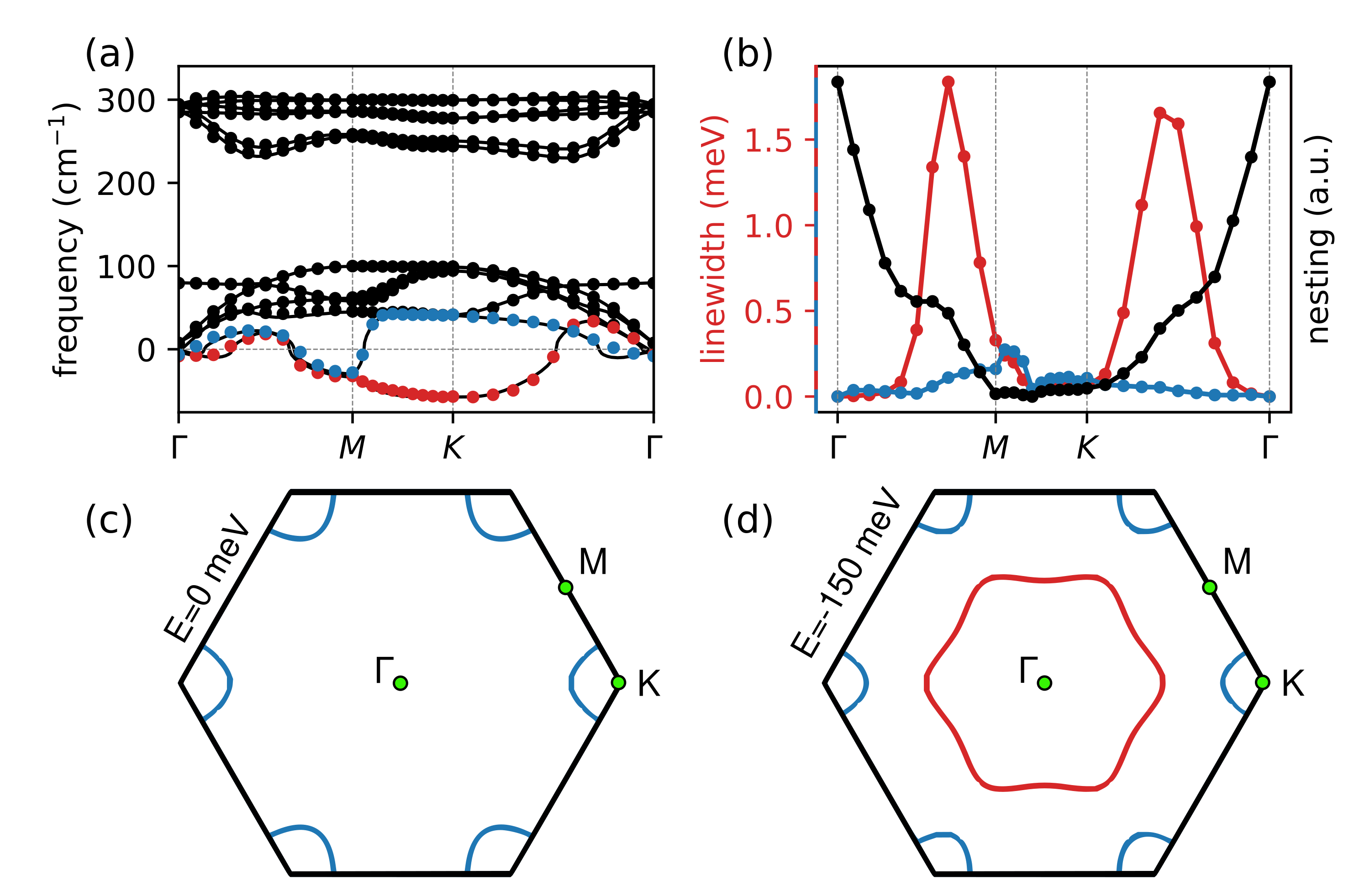}
\caption{\textbf{Electron-phonon linewidth, nesting function and Fermi surface of the SnP high-symmetry phase.}
	\textbf{a.} Phonon spectra of the SnP high-symmetry phase.
	The continuous line represents the interpolation derived from the 6x6x1 grid, while the dots signify specific DFPT calculations at individual $k$ points.
	\textbf{b.} Shown here are the electron-phonon linewidths for the two most unstable modes highlighted in blue/red in caption \textbf{a}.
	Additionally, the nesting function is represented along the $\mathrm{\Gamma M K \Gamma}$ path.
	\textbf{c.} The Fermi surface of the SnP high-symmetry phase exhibits a single Fermi pocket at the K point.
	\textbf{d.} Constant energy cut at an energy level of $-150$ meV.
	Besides the pocket at K, a new quasi-hexagonal pocket emerges, centered at $\Gamma$.
	This pocket harbors a high density of states due to the band flatness near $\Gamma$ (Fig. \ref{fig_monolayer}(e)), and it is responsible for the ``saddle-like" points in the nesting function (\textbf{b}) along the $\mathrm{\Gamma M}$ and $\mathrm{\Gamma K}$ lines.
	}
\label{fig:elph}
\end{figure}

We conducted computations of the electron-phonon linewidth and nesting function for the high-symmetry $\mathrm{SnP}$ phase.
As shown in Fig. \ref{fig:elph}(b), the electron-phonon linewidth displays notable peaks for the phonon branch responsible for the charge-density wave.
However, these peaks do not correspond to the most significant instabilities observed at the K point.
Moreover, neither the nesting function, depicted in Fig. \ref{fig:elph}(d), aligns with the trend of the unstable branch.
This discrepancy suggests that neither the nesting function nor the electron-phonon coupling serve as primary instigators of the observed CDW.
The contributions to the Born-Oppenheimer energy $V(R)$ can be decomposed into distinct components: $T_{e}+ V_{ee}+ V_{e-I}(R)+ V_{I-I}(R)$, where $R$ are the ionic positions and the terms represent the electronic kinetic energy, electron-electron interaction, electron-ion interaction and ion-ion interaction respectively.
While the electron-phonon coupling arises from the $V_{e-I}(R)$ term; in this instance, the minimization of the BO energy is predominantly dominated by the ion-ion interaction $V_{I-I}(R)$, leading to large structural changes between the different phases.
Consequently, it is not unexpected that both the nesting function and electron-phonon linewidth peak at different points than those of maximum instabilities.

\section{\label{app_structures}Crystal structures}
Table \ref{tab_Sn4P3} presents the Sn$_4$P$_3$ structure obtained from x-ray measurements, alongside the corresponding theoretical prediction. 
On the other hand, Table \ref{tab_monolayers} provides the theoretical predictions for the structural parameters of the individual layers comprising Sn$_4$P$_3$, namely SnP and Sn$_2$P. 
Additionally, the table also includes the CDW structures SnP-(K/M/$\Gamma$) resulting from the SnP monolayer.

\begin{table}[]
\begin{ruledtabular}
\begin{tabular}{cccc}
		\multicolumn{4}{c}{\textbf{Sn$_4$P$_3$ SG $R\overline{3}m$ (No. 166)}} \\ \hline

\multicolumn{2}{c}{Lattice parameter (\AA)} & theory & experiment \\ \hline
\multicolumn{2}{c}{$a=b$} & $4.003$ & $3.971$ \\
\multicolumn{2}{c}{$c$} & $35.798$ & $35.397$ \\ \hline \hline

atom & Wyckoff pos. & $\epsilon$ theory & $\epsilon$ experiment \\ \hline
P & \begin{tabular}[c]{@{}c@{}}$3a$\\ $(0,0,0)$\end{tabular} &  &  \\
P & \begin{tabular}[c]{@{}c@{}}$6c$\\ $(0,0,\pm \epsilon)$\end{tabular} & $0.42873$ & $0.42915$ \\
Sn & \begin{tabular}[c]{@{}c@{}}$6c$\\ $(0,0,\pm \epsilon)$\end{tabular} & $0.13339$ & $0.13406$ \\
Sn & \begin{tabular}[c]{@{}c@{}}$6c$\\ $(0,0,\pm \epsilon)$\end{tabular} & $0.28966$ & $0.28943$
\end{tabular}
\end{ruledtabular}
\caption{\label{tab_Sn4P3}
Experimental x-ray measurements and theoretical prediction of the Sn$_4$P$_3$ crystal structure.
}
\end{table}

\begin{table}[]
\begin{ruledtabular}
\begin{tabular}{ccccc}

\multicolumn{5}{c}{\textbf{SnP Monolayer SG $P\overline{3}m1$ (No. 164)}} \\ \hline
\multicolumn{3}{c}{$a=b=3.905$ \AA} & \multicolumn{2}{c}{$c=13.941$ \AA} \\ \hline
atom & Wyckoff pos. & $x$ & $y$ & $z$ \\ \hline
P & $2c$ & $0$ & $0$ & $0.18102$ \\
Sn & $2d$ & $1/3$ & $2/3$ & $0.09356$ \\ \hline \hline
\\

\multicolumn{5}{c}{\textbf{SnP-K Monolayer SG $P\overline{3}1m$ (No. 162)}} \\ \hline
\multicolumn{3}{c}{$a=b=6.779$ \AA} & \multicolumn{2}{c}{$c=25.157$ \AA} \\ \hline
atom & Wyckoff pos. & $x$ & $y$ & $z$ \\ \hline
P & $2e$ & $0$ & $0$ & $0.08316$ \\
P & $4h$ & $1/3$ & $2/3$ & $0.10945$ \\
Sn & $6k$ & $0.3700138$ & $0$ & $0.05352$ \\ \hline \hline
\\

\multicolumn{5}{c}{\textbf{SnP-M Monolayer SG $P2/m$ (No. 13)}} \\ \hline
\multicolumn{2}{c}{$\beta=0$} & $a=25.157$ \AA & $b=3.914$ \AA & $c=6.779$ \AA \\ \hline
atom & Wyckoff pos. & $x$ & $y$ & $z$ \\ \hline
P & $4g$ & $0.10091$ & $0.18215$ & $0.28476$ \\
Sn & $4g$ & $0.05363$ & $0.67507$ & $0.45998$ \\ \hline \hline
\\

\multicolumn{5}{c}{\textbf{SnP-$\Gamma$ Monolayer SG $P\overline{3}m1$ (No. 164)}} \\ \hline
\multicolumn{3}{c}{$a=b=3.960$ \AA} & \multicolumn{2}{c}{$c=25.157$ \AA} \\ \hline
atom & Wyckoff pos. & $x$ & $y$ & $z$ \\ \hline
P & $2d$ & $1/3$ & $2/3$ & $0.10334$ \\
Sn & $2c$ & $0$ & $0$ & $0.05726$ \\ \hline \hline
\\

\multicolumn{5}{c}{\textbf{Sn$_2$P Monolayer SG $P\overline{3}m1$ (No. 164)}} \\ \hline
\multicolumn{3}{c}{$a=b=3.684$ \AA} & \multicolumn{2}{c}{$c=16.000$ \AA} \\ \hline
atom & Wyckoff pos. & $x$ & $y$ & $z$ \\ \hline
P & $1a$ & $0$ & $0$ & $0$ \\
Sn & $2d$ & $1/3$ & $2/3$ & $0.10903$

\end{tabular}
\end{ruledtabular}
\caption{\label{tab_monolayers}
Calculated lattice parameters and atomic coordinates of all the studied monolayers.
SnP and Sn$_2$P are the two monolayers that originally form Sn$_4$P$_3$.
SnP-(K/M/$\Gamma$) are the resulting CDW structures after the condensation of the (K/M/$\Gamma$) unstable phonons of the SnP monolayer.
}
\end{table}

\FloatBarrier
\bibliography{bibtex.bib} %

\begin{thebibliography}{50}%
\makeatletter
\providecommand \@ifxundefined [1]{%
 \@ifx{#1\undefined}
}%
\providecommand \@ifnum [1]{%
 \ifnum #1\expandafter \@firstoftwo
 \else \expandafter \@secondoftwo
 \fi
}%
\providecommand \@ifx [1]{%
 \ifx #1\expandafter \@firstoftwo
 \else \expandafter \@secondoftwo
 \fi
}%
\providecommand \natexlab [1]{#1}%
\providecommand \enquote  [1]{``#1''}%
\providecommand \bibnamefont  [1]{#1}%
\providecommand \bibfnamefont [1]{#1}%
\providecommand \citenamefont [1]{#1}%
\providecommand \href@noop [0]{\@secondoftwo}%
\providecommand \href [0]{\begingroup \@sanitize@url \@href}%
\providecommand \@href[1]{\@@startlink{#1}\@@href}%
\providecommand \@@href[1]{\endgroup#1\@@endlink}%
\providecommand \@sanitize@url [0]{\catcode `\\12\catcode `\$12\catcode `\&12\catcode `\#12\catcode `\^12\catcode `\_12\catcode `\%12\relax}%
\providecommand \@@startlink[1]{}%
\providecommand \@@endlink[0]{}%
\providecommand \url  [0]{\begingroup\@sanitize@url \@url }%
\providecommand \@url [1]{\endgroup\@href {#1}{\urlprefix }}%
\providecommand \urlprefix  [0]{URL }%
\providecommand \Eprint [0]{\href }%
\providecommand \doibase [0]{https://doi.org/}%
\providecommand \selectlanguage [0]{\@gobble}%
\providecommand \bibinfo  [0]{\@secondoftwo}%
\providecommand \bibfield  [0]{\@secondoftwo}%
\providecommand \translation [1]{[#1]}%
\providecommand \BibitemOpen [0]{}%
\providecommand \bibitemStop [0]{}%
\providecommand \bibitemNoStop [0]{.\EOS\space}%
\providecommand \EOS [0]{\spacefactor3000\relax}%
\providecommand \BibitemShut  [1]{\csname bibitem#1\endcsname}%
\let\auto@bib@innerbib\@empty
%</preamble>
\bibitem [{\citenamefont {Grüner}(1988)}]{gruner_1988}%
  \BibitemOpen
  \bibfield  {author} {\bibinfo {author} {\bibfnamefont {G.}~\bibnamefont {Grüner}},\ }\bibfield  {title} {\bibinfo {title} {The dynamics of charge-density waves},\ }\href {https://doi.org/10.1103/RevModPhys.60.1129} {\bibfield  {journal} {\bibinfo  {journal} {Reviews of Modern Physics}\ }\textbf {\bibinfo {volume} {60}},\ \bibinfo {pages} {1129} (\bibinfo {year} {1988})}\BibitemShut {NoStop}%
\bibitem [{\citenamefont {Grüner}\ and\ \citenamefont {Zettl}(1985)}]{gruner_1985}%
  \BibitemOpen
  \bibfield  {author} {\bibinfo {author} {\bibfnamefont {G.}~\bibnamefont {Grüner}}\ and\ \bibinfo {author} {\bibfnamefont {A.}~\bibnamefont {Zettl}},\ }\bibfield  {title} {\bibinfo {title} {Charge density wave conduction: {{A}} novel collective transport phenomenon in solids},\ }\href {https://doi.org/10.1016/0370-1573(85)90073-0} {\bibfield  {journal} {\bibinfo  {journal} {Physics Reports}\ }\textbf {\bibinfo {volume} {119}},\ \bibinfo {pages} {117} (\bibinfo {year} {1985})}\BibitemShut {NoStop}%
\bibitem [{\citenamefont {Wang}\ \emph {et~al.}(1983)\citenamefont {Wang}, \citenamefont {{Saint-Lager}}, \citenamefont {Monceau}, \citenamefont {Renard}, \citenamefont {Gressier}, \citenamefont {Meerschaut}, \citenamefont {Guemas},\ and\ \citenamefont {Rouxel}}]{wang_1983}%
  \BibitemOpen
  \bibfield  {author} {\bibinfo {author} {\bibfnamefont {Z.~Z.}\ \bibnamefont {Wang}}, \bibinfo {author} {\bibfnamefont {M.~C.}\ \bibnamefont {{Saint-Lager}}}, \bibinfo {author} {\bibfnamefont {P.}~\bibnamefont {Monceau}}, \bibinfo {author} {\bibfnamefont {M.}~\bibnamefont {Renard}}, \bibinfo {author} {\bibfnamefont {P.}~\bibnamefont {Gressier}}, \bibinfo {author} {\bibfnamefont {A.}~\bibnamefont {Meerschaut}}, \bibinfo {author} {\bibfnamefont {L.}~\bibnamefont {Guemas}},\ and\ \bibinfo {author} {\bibfnamefont {J.}~\bibnamefont {Rouxel}},\ }\bibfield  {title} {\bibinfo {title} {Charge density wave transport in ({{TaSe4}}){{2I}}},\ }\href {https://doi.org/10.1016/0038-1098(83)90662-2} {\bibfield  {journal} {\bibinfo  {journal} {Solid State Communications}\ }\textbf {\bibinfo {volume} {46}},\ \bibinfo {pages} {325} (\bibinfo {year} {1983})}\BibitemShut {NoStop}%
\bibitem [{\citenamefont {Guo}\ \emph {et~al.}(2022)\citenamefont {Guo}, \citenamefont {Putzke}, \citenamefont {Konyzheva}, \citenamefont {Huang}, \citenamefont {{Gutierrez-Amigo}}, \citenamefont {Errea}, \citenamefont {Chen}, \citenamefont {Vergniory}, \citenamefont {Felser}, \citenamefont {Fischer}, \citenamefont {Neupert},\ and\ \citenamefont {Moll}}]{guo_2022}%
  \BibitemOpen
  \bibfield  {author} {\bibinfo {author} {\bibfnamefont {C.}~\bibnamefont {Guo}}, \bibinfo {author} {\bibfnamefont {C.}~\bibnamefont {Putzke}}, \bibinfo {author} {\bibfnamefont {S.}~\bibnamefont {Konyzheva}}, \bibinfo {author} {\bibfnamefont {X.}~\bibnamefont {Huang}}, \bibinfo {author} {\bibfnamefont {M.}~\bibnamefont {{Gutierrez-Amigo}}}, \bibinfo {author} {\bibfnamefont {I.}~\bibnamefont {Errea}}, \bibinfo {author} {\bibfnamefont {D.}~\bibnamefont {Chen}}, \bibinfo {author} {\bibfnamefont {M.~G.}\ \bibnamefont {Vergniory}}, \bibinfo {author} {\bibfnamefont {C.}~\bibnamefont {Felser}}, \bibinfo {author} {\bibfnamefont {M.~H.}\ \bibnamefont {Fischer}}, \bibinfo {author} {\bibfnamefont {T.}~\bibnamefont {Neupert}},\ and\ \bibinfo {author} {\bibfnamefont {P.~J.~W.}\ \bibnamefont {Moll}},\ }\bibfield  {title} {\bibinfo {title} {Switchable chiral transport in charge-ordered kagome metal {{CsV3Sb5}}},\ }\href {https://doi.org/10.1038/s41586-022-05127-9} {\bibfield  {journal} {\bibinfo  {journal} {Nature}\ ,\ \bibinfo {pages} {1}} (\bibinfo {year} {2022})}\BibitemShut {NoStop}%
\bibitem [{\citenamefont {Kwok}\ and\ \citenamefont {Brown}(1989)}]{kwok_1989}%
  \BibitemOpen
  \bibfield  {author} {\bibinfo {author} {\bibfnamefont {R.~S.}\ \bibnamefont {Kwok}}\ and\ \bibinfo {author} {\bibfnamefont {S.~E.}\ \bibnamefont {Brown}},\ }\bibfield  {title} {\bibinfo {title} {Thermal conductivity of the charge-density-wave systems {$\mathrm{K_{0.3}MoO_{3}}$} and {$\mathrm{(TaSe_{4})_{2}I}$} near the peierls transition},\ }\href {https://doi.org/10.1103/PhysRevLett.63.895} {\bibfield  {journal} {\bibinfo  {journal} {Physical Review Letters}\ }\textbf {\bibinfo {volume} {63}},\ \bibinfo {pages} {895} (\bibinfo {year} {1989})}\BibitemShut {NoStop}%
\bibitem [{\citenamefont {Kuo}\ \emph {et~al.}(2001)\citenamefont {Kuo}, \citenamefont {Lue}, \citenamefont {Hsu}, \citenamefont {Li},\ and\ \citenamefont {Yang}}]{kuo_2001}%
  \BibitemOpen
  \bibfield  {author} {\bibinfo {author} {\bibfnamefont {Y.-K.}\ \bibnamefont {Kuo}}, \bibinfo {author} {\bibfnamefont {C.~S.}\ \bibnamefont {Lue}}, \bibinfo {author} {\bibfnamefont {F.~H.}\ \bibnamefont {Hsu}}, \bibinfo {author} {\bibfnamefont {H.~H.}\ \bibnamefont {Li}},\ and\ \bibinfo {author} {\bibfnamefont {H.~D.}\ \bibnamefont {Yang}},\ }\bibfield  {title} {\bibinfo {title} {Thermal properties of {$\mathrm{Lu_{5}Ir_{4}Si_{10}}$} near the charge-density-wave transition},\ }\href {https://doi.org/10.1103/PhysRevB.64.125124} {\bibfield  {journal} {\bibinfo  {journal} {Physical Review B}\ }\textbf {\bibinfo {volume} {64}},\ \bibinfo {pages} {125124} (\bibinfo {year} {2001})}\BibitemShut {NoStop}%
\bibitem [{\citenamefont {Wang}\ \emph {et~al.}(2022)\citenamefont {Wang}, \citenamefont {Petrides}, \citenamefont {McNamara}, \citenamefont {Hosen}, \citenamefont {Lei}, \citenamefont {Wu}, \citenamefont {Hart}, \citenamefont {Lv}, \citenamefont {Yan}, \citenamefont {Xiao}, \citenamefont {Cha}, \citenamefont {Narang}, \citenamefont {Schoop},\ and\ \citenamefont {Burch}}]{wang_2022a}%
  \BibitemOpen
  \bibfield  {author} {\bibinfo {author} {\bibfnamefont {Y.}~\bibnamefont {Wang}}, \bibinfo {author} {\bibfnamefont {I.}~\bibnamefont {Petrides}}, \bibinfo {author} {\bibfnamefont {G.}~\bibnamefont {McNamara}}, \bibinfo {author} {\bibfnamefont {M.~M.}\ \bibnamefont {Hosen}}, \bibinfo {author} {\bibfnamefont {S.}~\bibnamefont {Lei}}, \bibinfo {author} {\bibfnamefont {Y.-C.}\ \bibnamefont {Wu}}, \bibinfo {author} {\bibfnamefont {J.~L.}\ \bibnamefont {Hart}}, \bibinfo {author} {\bibfnamefont {H.}~\bibnamefont {Lv}}, \bibinfo {author} {\bibfnamefont {J.}~\bibnamefont {Yan}}, \bibinfo {author} {\bibfnamefont {D.}~\bibnamefont {Xiao}}, \bibinfo {author} {\bibfnamefont {J.~J.}\ \bibnamefont {Cha}}, \bibinfo {author} {\bibfnamefont {P.}~\bibnamefont {Narang}}, \bibinfo {author} {\bibfnamefont {L.~M.}\ \bibnamefont {Schoop}},\ and\ \bibinfo {author} {\bibfnamefont {K.~S.}\ \bibnamefont {Burch}},\ }\bibfield  {title} {\bibinfo {title} {Axial {{Higgs}} mode detected by quantum pathway interference in {{RTe3}}},\ }\href {https://doi.org/10.1038/s41586-022-04746-6} {\bibfield  {journal} {\bibinfo  {journal} {Nature}\ }\textbf {\bibinfo {volume} {606}},\ \bibinfo {pages} {896} (\bibinfo {year} {2022})}\BibitemShut {NoStop}%
\bibitem [{\citenamefont {Gooth}\ \emph {et~al.}(2019)\citenamefont {Gooth}, \citenamefont {Bradlyn}, \citenamefont {Honnali}, \citenamefont {Schindler}, \citenamefont {Kumar}, \citenamefont {Noky}, \citenamefont {Qi}, \citenamefont {Shekhar}, \citenamefont {Sun}, \citenamefont {Wang}, \citenamefont {Bernevig},\ and\ \citenamefont {Felser}}]{gooth_2019}%
  \BibitemOpen
  \bibfield  {author} {\bibinfo {author} {\bibfnamefont {J.}~\bibnamefont {Gooth}}, \bibinfo {author} {\bibfnamefont {B.}~\bibnamefont {Bradlyn}}, \bibinfo {author} {\bibfnamefont {S.}~\bibnamefont {Honnali}}, \bibinfo {author} {\bibfnamefont {C.}~\bibnamefont {Schindler}}, \bibinfo {author} {\bibfnamefont {N.}~\bibnamefont {Kumar}}, \bibinfo {author} {\bibfnamefont {J.}~\bibnamefont {Noky}}, \bibinfo {author} {\bibfnamefont {Y.}~\bibnamefont {Qi}}, \bibinfo {author} {\bibfnamefont {C.}~\bibnamefont {Shekhar}}, \bibinfo {author} {\bibfnamefont {Y.}~\bibnamefont {Sun}}, \bibinfo {author} {\bibfnamefont {Z.}~\bibnamefont {Wang}}, \bibinfo {author} {\bibfnamefont {B.~A.}\ \bibnamefont {Bernevig}},\ and\ \bibinfo {author} {\bibfnamefont {C.}~\bibnamefont {Felser}},\ }\bibfield  {title} {\bibinfo {title} {Axionic charge-density wave in the {{Weyl}} semimetal ({{TaSe4}}){{2I}}},\ }\href {https://doi.org/10.1038/s41586-019-1630-4} {\bibfield  {journal} {\bibinfo  {journal} {Nature}\ }\textbf {\bibinfo {volume} {575}},\ \bibinfo {pages} {315} (\bibinfo {year} {2019})}\BibitemShut {NoStop}%
\bibitem [{\citenamefont {Wang}\ and\ \citenamefont {Zhang}(2013)}]{wang_2013}%
  \BibitemOpen
  \bibfield  {author} {\bibinfo {author} {\bibfnamefont {Z.}~\bibnamefont {Wang}}\ and\ \bibinfo {author} {\bibfnamefont {S.-C.}\ \bibnamefont {Zhang}},\ }\bibfield  {title} {\bibinfo {title} {Chiral anomaly, {{Charge Density Waves}}, and {{Axion Strings}} from {{Weyl Semimetals}}},\ }\href {https://doi.org/10.1103/PhysRevB.87.161107} {\bibfield  {journal} {\bibinfo  {journal} {Physical Review B}\ }\textbf {\bibinfo {volume} {87}},\ \bibinfo {pages} {161107} (\bibinfo {year} {2013})}\BibitemShut {NoStop}%
\bibitem [{\citenamefont {Wieder}\ \emph {et~al.}(2020)\citenamefont {Wieder}, \citenamefont {Lin},\ and\ \citenamefont {Bradlyn}}]{wieder_2020}%
  \BibitemOpen
  \bibfield  {author} {\bibinfo {author} {\bibfnamefont {B.~J.}\ \bibnamefont {Wieder}}, \bibinfo {author} {\bibfnamefont {K.-S.}\ \bibnamefont {Lin}},\ and\ \bibinfo {author} {\bibfnamefont {B.}~\bibnamefont {Bradlyn}},\ }\bibfield  {title} {\bibinfo {title} {Axionic band topology in inversion-symmetric {{Weyl-charge-density}} waves},\ }\href {https://doi.org/10.1103/PhysRevResearch.2.042010} {\bibfield  {journal} {\bibinfo  {journal} {Physical Review Research}\ }\textbf {\bibinfo {volume} {2}},\ \bibinfo {pages} {042010} (\bibinfo {year} {2020})}\BibitemShut {NoStop}%
\bibitem [{\citenamefont {Devescovi}\ \emph {et~al.}(2023)\citenamefont {Devescovi}, \citenamefont {{Morales-Pérez}}, \citenamefont {Hwang}, \citenamefont {{García-Díez}}, \citenamefont {Robredo}, \citenamefont {Mañes}, \citenamefont {Bradlyn}, \citenamefont {{García-Etxarri}},\ and\ \citenamefont {Vergniory}}]{devescovi_2023}%
  \BibitemOpen
  \bibfield  {author} {\bibinfo {author} {\bibfnamefont {C.}~\bibnamefont {Devescovi}}, \bibinfo {author} {\bibfnamefont {A.}~\bibnamefont {{Morales-Pérez}}}, \bibinfo {author} {\bibfnamefont {Y.}~\bibnamefont {Hwang}}, \bibinfo {author} {\bibfnamefont {M.}~\bibnamefont {{García-Díez}}}, \bibinfo {author} {\bibfnamefont {I.}~\bibnamefont {Robredo}}, \bibinfo {author} {\bibfnamefont {J.~L.}\ \bibnamefont {Mañes}}, \bibinfo {author} {\bibfnamefont {B.}~\bibnamefont {Bradlyn}}, \bibinfo {author} {\bibfnamefont {A.}~\bibnamefont {{García-Etxarri}}},\ and\ \bibinfo {author} {\bibfnamefont {M.~G.}\ \bibnamefont {Vergniory}},\ }\href {https://doi.org/10.48550/arXiv.2305.19805} {\bibinfo {title} {Axion {{Topology}} in {{Photonic Crystal Domain Walls}}}} (\bibinfo {year} {2023})\BibitemShut {NoStop}%
\bibitem [{\citenamefont {Lei}\ \emph {et~al.}(2021)\citenamefont {Lei}, \citenamefont {Teicher}, \citenamefont {Topp}, \citenamefont {Cai}, \citenamefont {Lin}, \citenamefont {Cheng}, \citenamefont {Salters}, \citenamefont {Rodolakis}, \citenamefont {McChesney}, \citenamefont {Lapidus}, \citenamefont {Yao}, \citenamefont {Krivenkov}, \citenamefont {Marchenko}, \citenamefont {Varykhalov}, \citenamefont {Ast}, \citenamefont {Car}, \citenamefont {Cano}, \citenamefont {Vergniory}, \citenamefont {Ong},\ and\ \citenamefont {Schoop}}]{lei_2021}%
  \BibitemOpen
  \bibfield  {author} {\bibinfo {author} {\bibfnamefont {S.}~\bibnamefont {Lei}}, \bibinfo {author} {\bibfnamefont {S.~M.~L.}\ \bibnamefont {Teicher}}, \bibinfo {author} {\bibfnamefont {A.}~\bibnamefont {Topp}}, \bibinfo {author} {\bibfnamefont {K.}~\bibnamefont {Cai}}, \bibinfo {author} {\bibfnamefont {J.}~\bibnamefont {Lin}}, \bibinfo {author} {\bibfnamefont {G.}~\bibnamefont {Cheng}}, \bibinfo {author} {\bibfnamefont {T.~H.}\ \bibnamefont {Salters}}, \bibinfo {author} {\bibfnamefont {F.}~\bibnamefont {Rodolakis}}, \bibinfo {author} {\bibfnamefont {J.~L.}\ \bibnamefont {McChesney}}, \bibinfo {author} {\bibfnamefont {S.}~\bibnamefont {Lapidus}}, \bibinfo {author} {\bibfnamefont {N.}~\bibnamefont {Yao}}, \bibinfo {author} {\bibfnamefont {M.}~\bibnamefont {Krivenkov}}, \bibinfo {author} {\bibfnamefont {D.}~\bibnamefont {Marchenko}}, \bibinfo {author} {\bibfnamefont {A.}~\bibnamefont {Varykhalov}}, \bibinfo {author} {\bibfnamefont {C.~R.}\ \bibnamefont {Ast}}, \bibinfo {author} {\bibfnamefont {R.}~\bibnamefont {Car}}, \bibinfo {author} {\bibfnamefont {J.}~\bibnamefont {Cano}}, \bibinfo {author} {\bibfnamefont {M.~G.}\ \bibnamefont {Vergniory}}, \bibinfo {author} {\bibfnamefont {N.~P.}\ \bibnamefont {Ong}},\ and\ \bibinfo {author} {\bibfnamefont {L.~M.}\ \bibnamefont {Schoop}},\ }\bibfield  {title} {\bibinfo {title} {Band {{Engineering}} of {{Dirac Semimetals Using Charge Density Waves}}},\ }\href {https://doi.org/10.1002/adma.202101591} {\bibfield  {journal} {\bibinfo  {journal} {Advanced Materials}\ }\textbf {\bibinfo {volume} {33}},\ \bibinfo {pages} {2101591} (\bibinfo {year} {2021})}\BibitemShut {NoStop}%
\bibitem [{\citenamefont {Weber}\ \emph {et~al.}(2011{\natexlab{a}})\citenamefont {Weber}, \citenamefont {Rosenkranz}, \citenamefont {Castellan}, \citenamefont {Osborn}, \citenamefont {Karapetrov}, \citenamefont {Hott}, \citenamefont {Heid}, \citenamefont {Bohnen},\ and\ \citenamefont {Alatas}}]{weber_2011a}%
  \BibitemOpen
  \bibfield  {author} {\bibinfo {author} {\bibfnamefont {F.}~\bibnamefont {Weber}}, \bibinfo {author} {\bibfnamefont {S.}~\bibnamefont {Rosenkranz}}, \bibinfo {author} {\bibfnamefont {J.-P.}\ \bibnamefont {Castellan}}, \bibinfo {author} {\bibfnamefont {R.}~\bibnamefont {Osborn}}, \bibinfo {author} {\bibfnamefont {G.}~\bibnamefont {Karapetrov}}, \bibinfo {author} {\bibfnamefont {R.}~\bibnamefont {Hott}}, \bibinfo {author} {\bibfnamefont {R.}~\bibnamefont {Heid}}, \bibinfo {author} {\bibfnamefont {K.-P.}\ \bibnamefont {Bohnen}},\ and\ \bibinfo {author} {\bibfnamefont {A.}~\bibnamefont {Alatas}},\ }\bibfield  {title} {\bibinfo {title} {Electron-phonon coupling and the soft phonon mode in {$\mathrm{TiSe_{2}}$}},\ }\href {https://doi.org/10.1103/PhysRevLett.107.266401} {\bibfield  {journal} {\bibinfo  {journal} {Physical Review Letters}\ }\textbf {\bibinfo {volume} {107}},\ \bibinfo {pages} {266401} (\bibinfo {year} {2011}{\natexlab{a}})}\BibitemShut {NoStop}%
\bibitem [{\citenamefont {Weber}\ \emph {et~al.}(2011{\natexlab{b}})\citenamefont {Weber}, \citenamefont {Rosenkranz}, \citenamefont {Castellan}, \citenamefont {Osborn}, \citenamefont {Hott}, \citenamefont {Heid}, \citenamefont {Bohnen}, \citenamefont {Egami}, \citenamefont {Said},\ and\ \citenamefont {Reznik}}]{weber_2011}%
  \BibitemOpen
  \bibfield  {author} {\bibinfo {author} {\bibfnamefont {F.}~\bibnamefont {Weber}}, \bibinfo {author} {\bibfnamefont {S.}~\bibnamefont {Rosenkranz}}, \bibinfo {author} {\bibfnamefont {J.-P.}\ \bibnamefont {Castellan}}, \bibinfo {author} {\bibfnamefont {R.}~\bibnamefont {Osborn}}, \bibinfo {author} {\bibfnamefont {R.}~\bibnamefont {Hott}}, \bibinfo {author} {\bibfnamefont {R.}~\bibnamefont {Heid}}, \bibinfo {author} {\bibfnamefont {K.-P.}\ \bibnamefont {Bohnen}}, \bibinfo {author} {\bibfnamefont {T.}~\bibnamefont {Egami}}, \bibinfo {author} {\bibfnamefont {A.~H.}\ \bibnamefont {Said}},\ and\ \bibinfo {author} {\bibfnamefont {D.}~\bibnamefont {Reznik}},\ }\bibfield  {title} {\bibinfo {title} {Extended phonon collapse and the origin of the charge-density wave in {$2H\mathrm{\text{\ensuremath{-}}NbSe_{2}}$}},\ }\href {https://doi.org/10.1103/PhysRevLett.107.107403} {\bibfield  {journal} {\bibinfo  {journal} {Physical Review Letters}\ }\textbf {\bibinfo {volume} {107}},\ \bibinfo {pages} {107403} (\bibinfo {year} {2011}{\natexlab{b}})}\BibitemShut {NoStop}%
\bibitem [{\citenamefont {Diego}\ \emph {et~al.}(2021)\citenamefont {Diego}, \citenamefont {Said}, \citenamefont {Mahatha}, \citenamefont {Bianco}, \citenamefont {Monacelli}, \citenamefont {Calandra}, \citenamefont {Mauri}, \citenamefont {Rossnagel}, \citenamefont {Errea},\ and\ \citenamefont {{Blanco-Canosa}}}]{diego_2021}%
  \BibitemOpen
  \bibfield  {author} {\bibinfo {author} {\bibfnamefont {J.}~\bibnamefont {Diego}}, \bibinfo {author} {\bibfnamefont {A.~H.}\ \bibnamefont {Said}}, \bibinfo {author} {\bibfnamefont {S.~K.}\ \bibnamefont {Mahatha}}, \bibinfo {author} {\bibfnamefont {R.}~\bibnamefont {Bianco}}, \bibinfo {author} {\bibfnamefont {L.}~\bibnamefont {Monacelli}}, \bibinfo {author} {\bibfnamefont {M.}~\bibnamefont {Calandra}}, \bibinfo {author} {\bibfnamefont {F.}~\bibnamefont {Mauri}}, \bibinfo {author} {\bibfnamefont {K.}~\bibnamefont {Rossnagel}}, \bibinfo {author} {\bibfnamefont {I.}~\bibnamefont {Errea}},\ and\ \bibinfo {author} {\bibfnamefont {S.}~\bibnamefont {{Blanco-Canosa}}},\ }\bibfield  {title} {\bibinfo {title} {Van der {{Waals}} driven anharmonic melting of the {{3D}} charge density wave in {{VSe2}}},\ }\href {https://doi.org/10.1038/s41467-020-20829-2} {\bibfield  {journal} {\bibinfo  {journal} {Nature Communications}\ }\textbf {\bibinfo {volume} {12}},\ \bibinfo {pages} {598} (\bibinfo {year} {2021})}\BibitemShut {NoStop}%
\bibitem [{\citenamefont {Zhou}\ \emph {et~al.}(2020{\natexlab{a}})\citenamefont {Zhou}, \citenamefont {Bianco}, \citenamefont {Monacelli}, \citenamefont {Errea}, \citenamefont {Mauri},\ and\ \citenamefont {Calandra}}]{zhou_2020a}%
  \BibitemOpen
  \bibfield  {author} {\bibinfo {author} {\bibfnamefont {J.~S.}\ \bibnamefont {Zhou}}, \bibinfo {author} {\bibfnamefont {R.}~\bibnamefont {Bianco}}, \bibinfo {author} {\bibfnamefont {L.}~\bibnamefont {Monacelli}}, \bibinfo {author} {\bibfnamefont {I.}~\bibnamefont {Errea}}, \bibinfo {author} {\bibfnamefont {F.}~\bibnamefont {Mauri}},\ and\ \bibinfo {author} {\bibfnamefont {M.}~\bibnamefont {Calandra}},\ }\bibfield  {title} {\bibinfo {title} {Theory of the thickness dependence of the charge density wave transition in 1 {{T-TiTe2}}},\ }\href {https://doi.org/10.1088/2053-1583/abae7a} {\bibfield  {journal} {\bibinfo  {journal} {2D Materials}\ }\textbf {\bibinfo {volume} {7}},\ \bibinfo {pages} {045032} (\bibinfo {year} {2020}{\natexlab{a}})}\BibitemShut {NoStop}%
\bibitem [{\citenamefont {Bianco}\ \emph {et~al.}(2020)\citenamefont {Bianco}, \citenamefont {Monacelli}, \citenamefont {Calandra}, \citenamefont {Mauri},\ and\ \citenamefont {Errea}}]{bianco_2020}%
  \BibitemOpen
  \bibfield  {author} {\bibinfo {author} {\bibfnamefont {R.}~\bibnamefont {Bianco}}, \bibinfo {author} {\bibfnamefont {L.}~\bibnamefont {Monacelli}}, \bibinfo {author} {\bibfnamefont {M.}~\bibnamefont {Calandra}}, \bibinfo {author} {\bibfnamefont {F.}~\bibnamefont {Mauri}},\ and\ \bibinfo {author} {\bibfnamefont {I.}~\bibnamefont {Errea}},\ }\bibfield  {title} {\bibinfo {title} {Weak dimensionality dependence and dominant role of ionic fluctuations in the charge-density-wave transition of {$\mathrm{NbSe_{2}}$}},\ }\href {https://doi.org/10.1103/PhysRevLett.125.106101} {\bibfield  {journal} {\bibinfo  {journal} {Physical Review Letters}\ }\textbf {\bibinfo {volume} {125}},\ \bibinfo {pages} {106101} (\bibinfo {year} {2020})}\BibitemShut {NoStop}%
\bibitem [{\citenamefont {Bianco}\ \emph {et~al.}(2019)\citenamefont {Bianco}, \citenamefont {Errea}, \citenamefont {Monacelli}, \citenamefont {Calandra},\ and\ \citenamefont {Mauri}}]{bianco_2019}%
  \BibitemOpen
  \bibfield  {author} {\bibinfo {author} {\bibfnamefont {R.}~\bibnamefont {Bianco}}, \bibinfo {author} {\bibfnamefont {I.}~\bibnamefont {Errea}}, \bibinfo {author} {\bibfnamefont {L.}~\bibnamefont {Monacelli}}, \bibinfo {author} {\bibfnamefont {M.}~\bibnamefont {Calandra}},\ and\ \bibinfo {author} {\bibfnamefont {F.}~\bibnamefont {Mauri}},\ }\bibfield  {title} {\bibinfo {title} {Quantum {{Enhancement}} of {{Charge Density Wave}} in {{NbS2}} in the {{Two-Dimensional Limit}}},\ }\href {https://doi.org/10.1021/acs.nanolett.9b00504} {\bibfield  {journal} {\bibinfo  {journal} {Nano Letters}\ }\textbf {\bibinfo {volume} {19}},\ \bibinfo {pages} {3098} (\bibinfo {year} {2019})}\BibitemShut {NoStop}%
\bibitem [{\citenamefont {Zhou}\ \emph {et~al.}(2020{\natexlab{b}})\citenamefont {Zhou}, \citenamefont {Monacelli}, \citenamefont {Bianco}, \citenamefont {Errea}, \citenamefont {Mauri},\ and\ \citenamefont {Calandra}}]{zhou_2020}%
  \BibitemOpen
  \bibfield  {author} {\bibinfo {author} {\bibfnamefont {J.~S.}\ \bibnamefont {Zhou}}, \bibinfo {author} {\bibfnamefont {L.}~\bibnamefont {Monacelli}}, \bibinfo {author} {\bibfnamefont {R.}~\bibnamefont {Bianco}}, \bibinfo {author} {\bibfnamefont {I.}~\bibnamefont {Errea}}, \bibinfo {author} {\bibfnamefont {F.}~\bibnamefont {Mauri}},\ and\ \bibinfo {author} {\bibfnamefont {M.}~\bibnamefont {Calandra}},\ }\bibfield  {title} {\bibinfo {title} {Anharmonicity and {{Doping Melt}} the {{Charge Density Wave}} in {{Single-Layer TiSe2}}},\ }\href {https://doi.org/10.1021/acs.nanolett.0c00597} {\bibfield  {journal} {\bibinfo  {journal} {Nano Letters}\ }\textbf {\bibinfo {volume} {20}},\ \bibinfo {pages} {4809} (\bibinfo {year} {2020}{\natexlab{b}})}\BibitemShut {NoStop}%
\bibitem [{\citenamefont {Fumega}\ \emph {et~al.}(2023)\citenamefont {Fumega}, \citenamefont {Diego}, \citenamefont {Pardo}, \citenamefont {{Blanco-Canosa}},\ and\ \citenamefont {Errea}}]{fumega_2023}%
  \BibitemOpen
  \bibfield  {author} {\bibinfo {author} {\bibfnamefont {A.~O.}\ \bibnamefont {Fumega}}, \bibinfo {author} {\bibfnamefont {J.}~\bibnamefont {Diego}}, \bibinfo {author} {\bibfnamefont {V.}~\bibnamefont {Pardo}}, \bibinfo {author} {\bibfnamefont {S.}~\bibnamefont {{Blanco-Canosa}}},\ and\ \bibinfo {author} {\bibfnamefont {I.}~\bibnamefont {Errea}},\ }\bibfield  {title} {\bibinfo {title} {Anharmonicity {{Reveals}} the {{Tunability}} of the {{Charge Density Wave Orders}} in {{Monolayer VSe2}}},\ }\href {https://doi.org/10.1021/acs.nanolett.2c04584} {\bibfield  {journal} {\bibinfo  {journal} {Nano Letters}\ }\textbf {\bibinfo {volume} {23}},\ \bibinfo {pages} {1794} (\bibinfo {year} {2023})}\BibitemShut {NoStop}%
\bibitem [{\citenamefont {Tallapally}\ \emph {et~al.}(2016)\citenamefont {Tallapally}, \citenamefont {Esteves}, \citenamefont {Nahar},\ and\ \citenamefont {Arachchige}}]{tallapally_2016}%
  \BibitemOpen
  \bibfield  {author} {\bibinfo {author} {\bibfnamefont {V.}~\bibnamefont {Tallapally}}, \bibinfo {author} {\bibfnamefont {R.~J.~A.}\ \bibnamefont {Esteves}}, \bibinfo {author} {\bibfnamefont {L.}~\bibnamefont {Nahar}},\ and\ \bibinfo {author} {\bibfnamefont {I.~U.}\ \bibnamefont {Arachchige}},\ }\bibfield  {title} {\bibinfo {title} {Multivariate {{Synthesis}} of {{Tin Phosphide Nanoparticles}}: {{Temperature}}, {{Time}}, and {{Ligand Control}} of {{Size}}, {{Shape}}, and {{Crystal Structure}}},\ }\href {https://doi.org/10.1021/acs.chemmater.6b01749} {\bibfield  {journal} {\bibinfo  {journal} {Chemistry of Materials}\ }\textbf {\bibinfo {volume} {28}},\ \bibinfo {pages} {5406} (\bibinfo {year} {2016})}\BibitemShut {NoStop}%
\bibitem [{\citenamefont {Li}\ \emph {et~al.}(2020)\citenamefont {Li}, \citenamefont {Shang}, \citenamefont {Zhao}, \citenamefont {Itkis}, \citenamefont {Jiao}, \citenamefont {Zhang}, \citenamefont {Liu},\ and\ \citenamefont {Song}}]{li_2020}%
  \BibitemOpen
  \bibfield  {author} {\bibinfo {author} {\bibfnamefont {B.}~\bibnamefont {Li}}, \bibinfo {author} {\bibfnamefont {S.}~\bibnamefont {Shang}}, \bibinfo {author} {\bibfnamefont {J.}~\bibnamefont {Zhao}}, \bibinfo {author} {\bibfnamefont {D.~M.}\ \bibnamefont {Itkis}}, \bibinfo {author} {\bibfnamefont {X.}~\bibnamefont {Jiao}}, \bibinfo {author} {\bibfnamefont {C.}~\bibnamefont {Zhang}}, \bibinfo {author} {\bibfnamefont {Z.-K.}\ \bibnamefont {Liu}},\ and\ \bibinfo {author} {\bibfnamefont {J.}~\bibnamefont {Song}},\ }\bibfield  {title} {\bibinfo {title} {Metastable trigonal {{SnP}}: {{A}} promising anode material for potassium-ion battery},\ }\href {https://doi.org/10.1016/j.carbon.2020.03.048} {\bibfield  {journal} {\bibinfo  {journal} {Carbon}\ }\textbf {\bibinfo {volume} {168}},\ \bibinfo {pages} {468} (\bibinfo {year} {2020})}\BibitemShut {NoStop}%
\bibitem [{\citenamefont {Vergniory}\ \emph {et~al.}(2022)\citenamefont {Vergniory}, \citenamefont {Wieder}, \citenamefont {Elcoro}, \citenamefont {Parkin}, \citenamefont {Felser}, \citenamefont {Bernevig},\ and\ \citenamefont {Regnault}}]{vergniory_2022}%
  \BibitemOpen
  \bibfield  {author} {\bibinfo {author} {\bibfnamefont {M.~G.}\ \bibnamefont {Vergniory}}, \bibinfo {author} {\bibfnamefont {B.~J.}\ \bibnamefont {Wieder}}, \bibinfo {author} {\bibfnamefont {L.}~\bibnamefont {Elcoro}}, \bibinfo {author} {\bibfnamefont {S.~S.~P.}\ \bibnamefont {Parkin}}, \bibinfo {author} {\bibfnamefont {C.}~\bibnamefont {Felser}}, \bibinfo {author} {\bibfnamefont {B.~A.}\ \bibnamefont {Bernevig}},\ and\ \bibinfo {author} {\bibfnamefont {N.}~\bibnamefont {Regnault}},\ }\bibfield  {title} {\bibinfo {title} {All topological bands of all nonmagnetic stoichiometric materials},\ }\href {https://doi.org/10.1126/science.abg9094} {\bibfield  {journal} {\bibinfo  {journal} {Science}\ }\textbf {\bibinfo {volume} {376}},\ \bibinfo {pages} {eabg9094} (\bibinfo {year} {2022})}\BibitemShut {NoStop}%
\bibitem [{\citenamefont {Vergniory}\ \emph {et~al.}(2019)\citenamefont {Vergniory}, \citenamefont {Elcoro}, \citenamefont {Felser}, \citenamefont {Regnault}, \citenamefont {Bernevig},\ and\ \citenamefont {Wang}}]{vergniory_2019}%
  \BibitemOpen
  \bibfield  {author} {\bibinfo {author} {\bibfnamefont {M.~G.}\ \bibnamefont {Vergniory}}, \bibinfo {author} {\bibfnamefont {L.}~\bibnamefont {Elcoro}}, \bibinfo {author} {\bibfnamefont {C.}~\bibnamefont {Felser}}, \bibinfo {author} {\bibfnamefont {N.}~\bibnamefont {Regnault}}, \bibinfo {author} {\bibfnamefont {B.~A.}\ \bibnamefont {Bernevig}},\ and\ \bibinfo {author} {\bibfnamefont {Z.}~\bibnamefont {Wang}},\ }\bibfield  {title} {\bibinfo {title} {A complete catalogue of high-quality topological materials},\ }\href {https://doi.org/10.1038/s41586-019-0954-4} {\bibfield  {journal} {\bibinfo  {journal} {Nature}\ }\textbf {\bibinfo {volume} {566}},\ \bibinfo {pages} {480} (\bibinfo {year} {2019})}\BibitemShut {NoStop}%
\bibitem [{\citenamefont {Bradlyn}\ \emph {et~al.}(2017)\citenamefont {Bradlyn}, \citenamefont {Elcoro}, \citenamefont {Cano}, \citenamefont {Vergniory}, \citenamefont {Wang}, \citenamefont {Felser}, \citenamefont {Aroyo},\ and\ \citenamefont {Bernevig}}]{bradlyn_2017}%
  \BibitemOpen
  \bibfield  {author} {\bibinfo {author} {\bibfnamefont {B.}~\bibnamefont {Bradlyn}}, \bibinfo {author} {\bibfnamefont {L.}~\bibnamefont {Elcoro}}, \bibinfo {author} {\bibfnamefont {J.}~\bibnamefont {Cano}}, \bibinfo {author} {\bibfnamefont {M.~G.}\ \bibnamefont {Vergniory}}, \bibinfo {author} {\bibfnamefont {Z.}~\bibnamefont {Wang}}, \bibinfo {author} {\bibfnamefont {C.}~\bibnamefont {Felser}}, \bibinfo {author} {\bibfnamefont {M.~I.}\ \bibnamefont {Aroyo}},\ and\ \bibinfo {author} {\bibfnamefont {B.~A.}\ \bibnamefont {Bernevig}},\ }\bibfield  {title} {\bibinfo {title} {Topological quantum chemistry},\ }\href {https://doi.org/10.1038/nature23268} {\bibfield  {journal} {\bibinfo  {journal} {Nature}\ }\textbf {\bibinfo {volume} {547}},\ \bibinfo {pages} {298} (\bibinfo {year} {2017})}\BibitemShut {NoStop}%
\bibitem [{\citenamefont {Errea}\ \emph {et~al.}(2014)\citenamefont {Errea}, \citenamefont {Calandra},\ and\ \citenamefont {Mauri}}]{errea_2014}%
  \BibitemOpen
  \bibfield  {author} {\bibinfo {author} {\bibfnamefont {I.}~\bibnamefont {Errea}}, \bibinfo {author} {\bibfnamefont {M.}~\bibnamefont {Calandra}},\ and\ \bibinfo {author} {\bibfnamefont {F.}~\bibnamefont {Mauri}},\ }\bibfield  {title} {\bibinfo {title} {Anharmonic free energies and phonon dispersions from the stochastic self-consistent harmonic approximation: {{Application}} to platinum and palladium hydrides},\ }\href {https://doi.org/10.1103/PhysRevB.89.064302} {\bibfield  {journal} {\bibinfo  {journal} {Physical Review B}\ }\textbf {\bibinfo {volume} {89}},\ \bibinfo {pages} {064302} (\bibinfo {year} {2014})}\BibitemShut {NoStop}%
\bibitem [{\citenamefont {Bianco}\ \emph {et~al.}(2017)\citenamefont {Bianco}, \citenamefont {Errea}, \citenamefont {Paulatto}, \citenamefont {Calandra},\ and\ \citenamefont {Mauri}}]{bianco_2017}%
  \BibitemOpen
  \bibfield  {author} {\bibinfo {author} {\bibfnamefont {R.}~\bibnamefont {Bianco}}, \bibinfo {author} {\bibfnamefont {I.}~\bibnamefont {Errea}}, \bibinfo {author} {\bibfnamefont {L.}~\bibnamefont {Paulatto}}, \bibinfo {author} {\bibfnamefont {M.}~\bibnamefont {Calandra}},\ and\ \bibinfo {author} {\bibfnamefont {F.}~\bibnamefont {Mauri}},\ }\bibfield  {title} {\bibinfo {title} {Second-order structural phase transitions, free energy curvature, and temperature-dependent anharmonic phonons in the self-consistent harmonic approximation: {{Theory}} and stochastic implementation},\ }\href {https://doi.org/10.1103/PhysRevB.96.014111} {\bibfield  {journal} {\bibinfo  {journal} {Physical Review B}\ }\textbf {\bibinfo {volume} {96}},\ \bibinfo {pages} {014111} (\bibinfo {year} {2017})}\BibitemShut {NoStop}%
\bibitem [{\citenamefont {Monacelli}\ \emph {et~al.}(2018)\citenamefont {Monacelli}, \citenamefont {Errea}, \citenamefont {Calandra},\ and\ \citenamefont {Mauri}}]{monacelli_2018}%
  \BibitemOpen
  \bibfield  {author} {\bibinfo {author} {\bibfnamefont {L.}~\bibnamefont {Monacelli}}, \bibinfo {author} {\bibfnamefont {I.}~\bibnamefont {Errea}}, \bibinfo {author} {\bibfnamefont {M.}~\bibnamefont {Calandra}},\ and\ \bibinfo {author} {\bibfnamefont {F.}~\bibnamefont {Mauri}},\ }\bibfield  {title} {\bibinfo {title} {Pressure and stress tensor of complex anharmonic crystals within the stochastic self-consistent harmonic approximation},\ }\href {https://doi.org/10.1103/PhysRevB.98.024106} {\bibfield  {journal} {\bibinfo  {journal} {Physical Review B}\ }\textbf {\bibinfo {volume} {98}},\ \bibinfo {pages} {024106} (\bibinfo {year} {2018})}\BibitemShut {NoStop}%
\bibitem [{\citenamefont {Monacelli}\ \emph {et~al.}(2021)\citenamefont {Monacelli}, \citenamefont {Bianco}, \citenamefont {Cherubini}, \citenamefont {Calandra}, \citenamefont {Errea},\ and\ \citenamefont {Mauri}}]{monacelli_2021}%
  \BibitemOpen
  \bibfield  {author} {\bibinfo {author} {\bibfnamefont {L.}~\bibnamefont {Monacelli}}, \bibinfo {author} {\bibfnamefont {R.}~\bibnamefont {Bianco}}, \bibinfo {author} {\bibfnamefont {M.}~\bibnamefont {Cherubini}}, \bibinfo {author} {\bibfnamefont {M.}~\bibnamefont {Calandra}}, \bibinfo {author} {\bibfnamefont {I.}~\bibnamefont {Errea}},\ and\ \bibinfo {author} {\bibfnamefont {F.}~\bibnamefont {Mauri}},\ }\bibfield  {title} {\bibinfo {title} {The stochastic self-consistent harmonic approximation: Calculating vibrational properties of materials with full quantum and anharmonic effects},\ }\href {https://doi.org/10.1088/1361-648X/ac066b} {\bibfield  {journal} {\bibinfo  {journal} {Journal of Physics: Condensed Matter}\ }\textbf {\bibinfo {volume} {33}},\ \bibinfo {pages} {363001} (\bibinfo {year} {2021})}\BibitemShut {NoStop}%
\bibitem [{\citenamefont {Olofsson}(1970)}]{olofsson_1970}%
  \BibitemOpen
  \bibfield  {author} {\bibinfo {author} {\bibfnamefont {O.}~\bibnamefont {Olofsson}},\ }\bibfield  {title} {\bibinfo {title} {X-{{Ray Investigations}} of the {{Tin-Phosphorus System}}.},\ }\href {https://doi.org/10.3891/acta.chem.scand.24-1153} {\bibfield  {journal} {\bibinfo  {journal} {Acta Chemica Scandinavica}\ }\textbf {\bibinfo {volume} {24}},\ \bibinfo {pages} {1153} (\bibinfo {year} {1970})}\BibitemShut {NoStop}%
\bibitem [{\citenamefont {Mounet}\ \emph {et~al.}(2018)\citenamefont {Mounet}, \citenamefont {Gibertini}, \citenamefont {Schwaller}, \citenamefont {Campi}, \citenamefont {Merkys}, \citenamefont {Marrazzo}, \citenamefont {Sohier}, \citenamefont {Castelli}, \citenamefont {Cepellotti}, \citenamefont {Pizzi},\ and\ \citenamefont {Marzari}}]{mounet_2018}%
  \BibitemOpen
  \bibfield  {author} {\bibinfo {author} {\bibfnamefont {N.}~\bibnamefont {Mounet}}, \bibinfo {author} {\bibfnamefont {M.}~\bibnamefont {Gibertini}}, \bibinfo {author} {\bibfnamefont {P.}~\bibnamefont {Schwaller}}, \bibinfo {author} {\bibfnamefont {D.}~\bibnamefont {Campi}}, \bibinfo {author} {\bibfnamefont {A.}~\bibnamefont {Merkys}}, \bibinfo {author} {\bibfnamefont {A.}~\bibnamefont {Marrazzo}}, \bibinfo {author} {\bibfnamefont {T.}~\bibnamefont {Sohier}}, \bibinfo {author} {\bibfnamefont {I.~E.}\ \bibnamefont {Castelli}}, \bibinfo {author} {\bibfnamefont {A.}~\bibnamefont {Cepellotti}}, \bibinfo {author} {\bibfnamefont {G.}~\bibnamefont {Pizzi}},\ and\ \bibinfo {author} {\bibfnamefont {N.}~\bibnamefont {Marzari}},\ }\bibfield  {title} {\bibinfo {title} {Two-dimensional materials from high-throughput computational exfoliation of experimentally known compounds},\ }\href {https://doi.org/10.1038/s41565-017-0035-5} {\bibfield  {journal} {\bibinfo  {journal} {Nature Nanotechnology}\ }\textbf {\bibinfo {volume} {13}},\ \bibinfo {pages} {246} (\bibinfo {year} {2018})}\BibitemShut {NoStop}%
\bibitem [{\citenamefont {Campi}\ \emph {et~al.}(2023)\citenamefont {Campi}, \citenamefont {Mounet}, \citenamefont {Gibertini}, \citenamefont {Pizzi},\ and\ \citenamefont {Marzari}}]{campi_2023}%
  \BibitemOpen
  \bibfield  {author} {\bibinfo {author} {\bibfnamefont {D.}~\bibnamefont {Campi}}, \bibinfo {author} {\bibfnamefont {N.}~\bibnamefont {Mounet}}, \bibinfo {author} {\bibfnamefont {M.}~\bibnamefont {Gibertini}}, \bibinfo {author} {\bibfnamefont {G.}~\bibnamefont {Pizzi}},\ and\ \bibinfo {author} {\bibfnamefont {N.}~\bibnamefont {Marzari}},\ }\bibfield  {title} {\bibinfo {title} {Expansion of the {{Materials Cloud 2D Database}}},\ }\href {https://doi.org/10.1021/acsnano.2c11510} {\bibfield  {journal} {\bibinfo  {journal} {ACS Nano}\ }\textbf {\bibinfo {volume} {17}},\ \bibinfo {pages} {11268} (\bibinfo {year} {2023})}\BibitemShut {NoStop}%
\bibitem [{\citenamefont {Chang}\ \emph {et~al.}(2016)\citenamefont {Chang}, \citenamefont {Liu}, \citenamefont {Lin}, \citenamefont {Wang}, \citenamefont {Zhao}, \citenamefont {Zhang}, \citenamefont {Jin}, \citenamefont {Zhong}, \citenamefont {Hu}, \citenamefont {Duan}, \citenamefont {Zhang}, \citenamefont {Fu}, \citenamefont {Xue}, \citenamefont {Chen},\ and\ \citenamefont {Ji}}]{chang_2016}%
  \BibitemOpen
  \bibfield  {author} {\bibinfo {author} {\bibfnamefont {K.}~\bibnamefont {Chang}}, \bibinfo {author} {\bibfnamefont {J.}~\bibnamefont {Liu}}, \bibinfo {author} {\bibfnamefont {H.}~\bibnamefont {Lin}}, \bibinfo {author} {\bibfnamefont {N.}~\bibnamefont {Wang}}, \bibinfo {author} {\bibfnamefont {K.}~\bibnamefont {Zhao}}, \bibinfo {author} {\bibfnamefont {A.}~\bibnamefont {Zhang}}, \bibinfo {author} {\bibfnamefont {F.}~\bibnamefont {Jin}}, \bibinfo {author} {\bibfnamefont {Y.}~\bibnamefont {Zhong}}, \bibinfo {author} {\bibfnamefont {X.}~\bibnamefont {Hu}}, \bibinfo {author} {\bibfnamefont {W.}~\bibnamefont {Duan}}, \bibinfo {author} {\bibfnamefont {Q.}~\bibnamefont {Zhang}}, \bibinfo {author} {\bibfnamefont {L.}~\bibnamefont {Fu}}, \bibinfo {author} {\bibfnamefont {Q.-K.}\ \bibnamefont {Xue}}, \bibinfo {author} {\bibfnamefont {X.}~\bibnamefont {Chen}},\ and\ \bibinfo {author} {\bibfnamefont {S.-H.}\ \bibnamefont {Ji}},\ }\bibfield  {title} {\bibinfo {title} {Discovery of robust in-plane ferroelectricity in atomic-thick {{SnTe}}},\ }\href {https://doi.org/10.1126/science.aad8609} {\bibfield  {journal} {\bibinfo  {journal} {Science}\ }\textbf {\bibinfo {volume} {353}},\ \bibinfo {pages} {274} (\bibinfo {year} {2016})}\BibitemShut {NoStop}%
\bibitem [{\citenamefont {Sasaki}\ \emph {et~al.}(1996)\citenamefont {Sasaki}, \citenamefont {Watanabe}, \citenamefont {Hashizume}, \citenamefont {Yamada},\ and\ \citenamefont {Nakazawa}}]{sasaki_1996}%
  \BibitemOpen
  \bibfield  {author} {\bibinfo {author} {\bibfnamefont {T.}~\bibnamefont {Sasaki}}, \bibinfo {author} {\bibfnamefont {M.}~\bibnamefont {Watanabe}}, \bibinfo {author} {\bibfnamefont {H.}~\bibnamefont {Hashizume}}, \bibinfo {author} {\bibfnamefont {H.}~\bibnamefont {Yamada}},\ and\ \bibinfo {author} {\bibfnamefont {H.}~\bibnamefont {Nakazawa}},\ }\bibfield  {title} {\bibinfo {title} {Macromolecule-like {{Aspects}} for a {{Colloidal Suspension}} of an {{Exfoliated Titanate}}. {{Pairwise Association}} of {{Nanosheets}} and {{Dynamic Reassembling Process Initiated}} from {{It}}},\ }\href {https://doi.org/10.1021/ja960073b} {\bibfield  {journal} {\bibinfo  {journal} {Journal of the American Chemical Society}\ }\textbf {\bibinfo {volume} {118}},\ \bibinfo {pages} {8329} (\bibinfo {year} {1996})}\BibitemShut {NoStop}%
\bibitem [{\citenamefont {Yuan}\ \emph {et~al.}(2022)\citenamefont {Yuan}, \citenamefont {Song}, \citenamefont {Cheng}, \citenamefont {Yao}, \citenamefont {Mozharivskyj},\ and\ \citenamefont {Schoop}}]{yuan_2022}%
  \BibitemOpen
  \bibfield  {author} {\bibinfo {author} {\bibfnamefont {F.}~\bibnamefont {Yuan}}, \bibinfo {author} {\bibfnamefont {X.}~\bibnamefont {Song}}, \bibinfo {author} {\bibfnamefont {G.}~\bibnamefont {Cheng}}, \bibinfo {author} {\bibfnamefont {N.}~\bibnamefont {Yao}}, \bibinfo {author} {\bibfnamefont {Y.}~\bibnamefont {Mozharivskyj}},\ and\ \bibinfo {author} {\bibfnamefont {L.~M.}\ \bibnamefont {Schoop}},\ }\bibfield  {title} {\bibinfo {title} {{Magnetic Nanosheets via Chemical Exfoliation of K2 xMnxSn1-xS2}},\ }\href {https://doi.org/10.1021/acs.chemmater.2c00488} {\bibfield  {journal} {\bibinfo  {journal} {Chemistry of Materials}\ }\textbf {\bibinfo {volume} {34}},\ \bibinfo {pages} {5084} (\bibinfo {year} {2022})}\BibitemShut {NoStop}%
\bibitem [{\citenamefont {Yang}\ \emph {et~al.}(2020)\citenamefont {Yang}, \citenamefont {Hong}, \citenamefont {Hao}, \citenamefont {Zhang}, \citenamefont {Liang}, \citenamefont {Long}, \citenamefont {Liu}, \citenamefont {Liu}, \citenamefont {Pang}, \citenamefont {Chen},\ and\ \citenamefont {Guo}}]{yang_2020}%
  \BibitemOpen
  \bibfield  {author} {\bibinfo {author} {\bibfnamefont {F.}~\bibnamefont {Yang}}, \bibinfo {author} {\bibfnamefont {J.}~\bibnamefont {Hong}}, \bibinfo {author} {\bibfnamefont {J.}~\bibnamefont {Hao}}, \bibinfo {author} {\bibfnamefont {S.}~\bibnamefont {Zhang}}, \bibinfo {author} {\bibfnamefont {G.}~\bibnamefont {Liang}}, \bibinfo {author} {\bibfnamefont {J.}~\bibnamefont {Long}}, \bibinfo {author} {\bibfnamefont {Y.}~\bibnamefont {Liu}}, \bibinfo {author} {\bibfnamefont {N.}~\bibnamefont {Liu}}, \bibinfo {author} {\bibfnamefont {W.~K.}\ \bibnamefont {Pang}}, \bibinfo {author} {\bibfnamefont {J.}~\bibnamefont {Chen}},\ and\ \bibinfo {author} {\bibfnamefont {Z.}~\bibnamefont {Guo}},\ }\bibfield  {title} {\bibinfo {title} {Ultrathin {{Few-Layer GeP Nanosheets}} via {{Lithiation-Assisted Chemical Exfoliation}} and {{Their Application}} in {{Sodium Storage}}},\ }\href {https://doi.org/10.1002/aenm.201903826} {\bibfield  {journal} {\bibinfo  {journal} {Advanced Energy Materials}\ }\textbf {\bibinfo {volume} {10}},\ \bibinfo {pages} {1903826} (\bibinfo {year} {2020})}\BibitemShut {NoStop}%
\bibitem [{\citenamefont {Fasolino}\ \emph {et~al.}(2007)\citenamefont {Fasolino}, \citenamefont {Los},\ and\ \citenamefont {Katsnelson}}]{fasolino_2007}%
  \BibitemOpen
  \bibfield  {author} {\bibinfo {author} {\bibfnamefont {A.}~\bibnamefont {Fasolino}}, \bibinfo {author} {\bibfnamefont {J.~H.}\ \bibnamefont {Los}},\ and\ \bibinfo {author} {\bibfnamefont {M.~I.}\ \bibnamefont {Katsnelson}},\ }\bibfield  {title} {\bibinfo {title} {Intrinsic ripples in graphene},\ }\href {https://doi.org/10.1038/nmat2011} {\bibfield  {journal} {\bibinfo  {journal} {Nature Materials}\ }\textbf {\bibinfo {volume} {6}},\ \bibinfo {pages} {858} (\bibinfo {year} {2007})}\BibitemShut {NoStop}%
\bibitem [{\citenamefont {Stokes}\ and\ \citenamefont {Hatch}(1988)}]{stokes_1988}%
  \BibitemOpen
  \bibfield  {author} {\bibinfo {author} {\bibfnamefont {H.~T.}\ \bibnamefont {Stokes}}\ and\ \bibinfo {author} {\bibfnamefont {D.~M.}\ \bibnamefont {Hatch}},\ }\href@noop {} {\emph {\bibinfo {title} {Isotropy Subgroups of the 230 Crystallographic Space Groups}}}\ (\bibinfo  {publisher} {World Scientific},\ \bibinfo {address} {Singapore ; Teaneck, NY, USA},\ \bibinfo {year} {1988})\BibitemShut {NoStop}%
\bibitem [{\citenamefont {Heyd}\ \emph {et~al.}(2003)\citenamefont {Heyd}, \citenamefont {Scuseria},\ and\ \citenamefont {Ernzerhof}}]{heyd_2003}%
  \BibitemOpen
  \bibfield  {author} {\bibinfo {author} {\bibfnamefont {J.}~\bibnamefont {Heyd}}, \bibinfo {author} {\bibfnamefont {G.~E.}\ \bibnamefont {Scuseria}},\ and\ \bibinfo {author} {\bibfnamefont {M.}~\bibnamefont {Ernzerhof}},\ }\bibfield  {title} {\bibinfo {title} {Hybrid functionals based on a screened {{Coulomb}} potential},\ }\href {https://doi.org/10.1063/1.1564060} {\bibfield  {journal} {\bibinfo  {journal} {The Journal of Chemical Physics}\ }\textbf {\bibinfo {volume} {118}},\ \bibinfo {pages} {8207} (\bibinfo {year} {2003})}\BibitemShut {NoStop}%
\bibitem [{\citenamefont {Lee}\ \emph {et~al.}(2008)\citenamefont {Lee}, \citenamefont {Wei}, \citenamefont {Kysar},\ and\ \citenamefont {Hone}}]{lee_2008}%
  \BibitemOpen
  \bibfield  {author} {\bibinfo {author} {\bibfnamefont {C.}~\bibnamefont {Lee}}, \bibinfo {author} {\bibfnamefont {X.}~\bibnamefont {Wei}}, \bibinfo {author} {\bibfnamefont {J.~W.}\ \bibnamefont {Kysar}},\ and\ \bibinfo {author} {\bibfnamefont {J.}~\bibnamefont {Hone}},\ }\bibfield  {title} {\bibinfo {title} {Measurement of the {{Elastic Properties}} and {{Intrinsic Strength}} of {{Monolayer Graphene}}},\ }\href {https://doi.org/10.1126/science.1157996} {\bibfield  {journal} {\bibinfo  {journal} {Science}\ }\textbf {\bibinfo {volume} {321}},\ \bibinfo {pages} {385} (\bibinfo {year} {2008})}\BibitemShut {NoStop}%
\bibitem [{\citenamefont {Bertolazzi}\ \emph {et~al.}(2011)\citenamefont {Bertolazzi}, \citenamefont {Brivio},\ and\ \citenamefont {Kis}}]{bertolazzi_2011}%
  \BibitemOpen
  \bibfield  {author} {\bibinfo {author} {\bibfnamefont {S.}~\bibnamefont {Bertolazzi}}, \bibinfo {author} {\bibfnamefont {J.}~\bibnamefont {Brivio}},\ and\ \bibinfo {author} {\bibfnamefont {A.}~\bibnamefont {Kis}},\ }\bibfield  {title} {\bibinfo {title} {Stretching and {{Breaking}} of {{Ultrathin MoS2}}},\ }\href {https://doi.org/10.1021/nn203879f} {\bibfield  {journal} {\bibinfo  {journal} {ACS Nano}\ }\textbf {\bibinfo {volume} {5}},\ \bibinfo {pages} {9703} (\bibinfo {year} {2011})}\BibitemShut {NoStop}%
\bibitem [{\citenamefont {Giannozzi}\ \emph {et~al.}(2009)\citenamefont {Giannozzi}, \citenamefont {Baroni}, \citenamefont {Bonini}, \citenamefont {Calandra}, \citenamefont {Car}, \citenamefont {Cavazzoni}, \citenamefont {Ceresoli}, \citenamefont {Chiarotti}, \citenamefont {Cococcioni}, \citenamefont {Dabo}, \citenamefont {Corso}, \citenamefont {de~Gironcoli}, \citenamefont {Fabris}, \citenamefont {Fratesi}, \citenamefont {Gebauer}, \citenamefont {Gerstmann}, \citenamefont {Gougoussis}, \citenamefont {Kokalj}, \citenamefont {Lazzeri}, \citenamefont {{Martin-Samos}}, \citenamefont {Marzari}, \citenamefont {Mauri}, \citenamefont {Mazzarello}, \citenamefont {Paolini}, \citenamefont {Pasquarello}, \citenamefont {Paulatto}, \citenamefont {Sbraccia}, \citenamefont {Scandolo}, \citenamefont {Sclauzero}, \citenamefont {Seitsonen}, \citenamefont {Smogunov}, \citenamefont {Umari},\ and\ \citenamefont {Wentzcovitch}}]{giannozzi_2009}%
  \BibitemOpen
  \bibfield  {author} {\bibinfo {author} {\bibfnamefont {P.}~\bibnamefont {Giannozzi}}, \bibinfo {author} {\bibfnamefont {S.}~\bibnamefont {Baroni}}, \bibinfo {author} {\bibfnamefont {N.}~\bibnamefont {Bonini}}, \bibinfo {author} {\bibfnamefont {M.}~\bibnamefont {Calandra}}, \bibinfo {author} {\bibfnamefont {R.}~\bibnamefont {Car}}, \bibinfo {author} {\bibfnamefont {C.}~\bibnamefont {Cavazzoni}}, \bibinfo {author} {\bibfnamefont {D.}~\bibnamefont {Ceresoli}}, \bibinfo {author} {\bibfnamefont {G.~L.}\ \bibnamefont {Chiarotti}}, \bibinfo {author} {\bibfnamefont {M.}~\bibnamefont {Cococcioni}}, \bibinfo {author} {\bibfnamefont {I.}~\bibnamefont {Dabo}}, \bibinfo {author} {\bibfnamefont {A.~D.}\ \bibnamefont {Corso}}, \bibinfo {author} {\bibfnamefont {S.}~\bibnamefont {de~Gironcoli}}, \bibinfo {author} {\bibfnamefont {S.}~\bibnamefont {Fabris}}, \bibinfo {author} {\bibfnamefont {G.}~\bibnamefont {Fratesi}}, \bibinfo {author} {\bibfnamefont {R.}~\bibnamefont {Gebauer}}, \bibinfo {author} {\bibfnamefont {U.}~\bibnamefont {Gerstmann}}, \bibinfo {author} {\bibfnamefont {C.}~\bibnamefont {Gougoussis}}, \bibinfo {author} {\bibfnamefont {A.}~\bibnamefont {Kokalj}}, \bibinfo {author} {\bibfnamefont {M.}~\bibnamefont {Lazzeri}}, \bibinfo {author} {\bibfnamefont {L.}~\bibnamefont {{Martin-Samos}}}, \bibinfo {author} {\bibfnamefont {N.}~\bibnamefont {Marzari}}, \bibinfo {author} {\bibfnamefont {F.}~\bibnamefont {Mauri}}, \bibinfo {author} {\bibfnamefont {R.}~\bibnamefont {Mazzarello}}, \bibinfo {author} {\bibfnamefont {S.}~\bibnamefont {Paolini}}, \bibinfo {author} {\bibfnamefont {A.}~\bibnamefont {Pasquarello}}, \bibinfo {author} {\bibfnamefont {L.}~\bibnamefont {Paulatto}}, \bibinfo {author} {\bibfnamefont {C.}~\bibnamefont {Sbraccia}}, \bibinfo {author} {\bibfnamefont {S.}~\bibnamefont {Scandolo}}, \bibinfo {author} {\bibfnamefont {G.}~\bibnamefont {Sclauzero}}, \bibinfo {author} {\bibfnamefont {A.~P.}\ \bibnamefont {Seitsonen}}, \bibinfo {author} {\bibfnamefont
  {A.}~\bibnamefont {Smogunov}}, \bibinfo {author} {\bibfnamefont {P.}~\bibnamefont {Umari}},\ and\ \bibinfo {author} {\bibfnamefont {R.~M.}\ \bibnamefont {Wentzcovitch}},\ }\bibfield  {title} {\bibinfo {title} {{{QUANTUM ESPRESSO}}: A modular and open-source software project for quantum simulations of materials},\ }\href {https://doi.org/10.1088/0953-8984/21/39/395502} {\bibfield  {journal} {\bibinfo  {journal} {Journal of Physics: Condensed Matter}\ }\textbf {\bibinfo {volume} {21}},\ \bibinfo {pages} {395502} (\bibinfo {year} {2009})}\BibitemShut {NoStop}%
\bibitem [{\citenamefont {Giannozzi}\ \emph {et~al.}(2017)\citenamefont {Giannozzi}, \citenamefont {Andreussi}, \citenamefont {Brumme}, \citenamefont {Bunau}, \citenamefont {Nardelli}, \citenamefont {Calandra}, \citenamefont {Car}, \citenamefont {Cavazzoni}, \citenamefont {Ceresoli}, \citenamefont {Cococcioni}, \citenamefont {Colonna}, \citenamefont {Carnimeo}, \citenamefont {Corso}, \citenamefont {de~Gironcoli}, \citenamefont {Delugas}, \citenamefont {DiStasio}, \citenamefont {Ferretti}, \citenamefont {Floris}, \citenamefont {Fratesi}, \citenamefont {Fugallo}, \citenamefont {Gebauer}, \citenamefont {Gerstmann}, \citenamefont {Giustino}, \citenamefont {Gorni}, \citenamefont {Jia}, \citenamefont {Kawamura}, \citenamefont {Ko}, \citenamefont {Kokalj}, \citenamefont {Küçükbenli}, \citenamefont {Lazzeri}, \citenamefont {Marsili}, \citenamefont {Marzari}, \citenamefont {Mauri}, \citenamefont {Nguyen}, \citenamefont {Nguyen}, \citenamefont {{Otero-de-la-Roza}}, \citenamefont {Paulatto}, \citenamefont {Poncé}, \citenamefont {Rocca}, \citenamefont {Sabatini}, \citenamefont {Santra}, \citenamefont {Schlipf}, \citenamefont {Seitsonen}, \citenamefont {Smogunov}, \citenamefont {Timrov}, \citenamefont {Thonhauser}, \citenamefont {Umari}, \citenamefont {Vast}, \citenamefont {Wu},\ and\ \citenamefont {Baroni}}]{giannozzi_2017}%
  \BibitemOpen
  \bibfield  {author} {\bibinfo {author} {\bibfnamefont {P.}~\bibnamefont {Giannozzi}}, \bibinfo {author} {\bibfnamefont {O.}~\bibnamefont {Andreussi}}, \bibinfo {author} {\bibfnamefont {T.}~\bibnamefont {Brumme}}, \bibinfo {author} {\bibfnamefont {O.}~\bibnamefont {Bunau}}, \bibinfo {author} {\bibfnamefont {M.~B.}\ \bibnamefont {Nardelli}}, \bibinfo {author} {\bibfnamefont {M.}~\bibnamefont {Calandra}}, \bibinfo {author} {\bibfnamefont {R.}~\bibnamefont {Car}}, \bibinfo {author} {\bibfnamefont {C.}~\bibnamefont {Cavazzoni}}, \bibinfo {author} {\bibfnamefont {D.}~\bibnamefont {Ceresoli}}, \bibinfo {author} {\bibfnamefont {M.}~\bibnamefont {Cococcioni}}, \bibinfo {author} {\bibfnamefont {N.}~\bibnamefont {Colonna}}, \bibinfo {author} {\bibfnamefont {I.}~\bibnamefont {Carnimeo}}, \bibinfo {author} {\bibfnamefont {A.~D.}\ \bibnamefont {Corso}}, \bibinfo {author} {\bibfnamefont {S.}~\bibnamefont {de~Gironcoli}}, \bibinfo {author} {\bibfnamefont {P.}~\bibnamefont {Delugas}}, \bibinfo {author} {\bibfnamefont {R.~A.}\ \bibnamefont {DiStasio}}, \bibinfo {author} {\bibfnamefont {A.}~\bibnamefont {Ferretti}}, \bibinfo {author} {\bibfnamefont {A.}~\bibnamefont {Floris}}, \bibinfo {author} {\bibfnamefont {G.}~\bibnamefont {Fratesi}}, \bibinfo {author} {\bibfnamefont {G.}~\bibnamefont {Fugallo}}, \bibinfo {author} {\bibfnamefont {R.}~\bibnamefont {Gebauer}}, \bibinfo {author} {\bibfnamefont {U.}~\bibnamefont {Gerstmann}}, \bibinfo {author} {\bibfnamefont {F.}~\bibnamefont {Giustino}}, \bibinfo {author} {\bibfnamefont {T.}~\bibnamefont {Gorni}}, \bibinfo {author} {\bibfnamefont {J.}~\bibnamefont {Jia}}, \bibinfo {author} {\bibfnamefont {M.}~\bibnamefont {Kawamura}}, \bibinfo {author} {\bibfnamefont {H.-Y.}\ \bibnamefont {Ko}}, \bibinfo {author} {\bibfnamefont {A.}~\bibnamefont {Kokalj}}, \bibinfo {author} {\bibfnamefont {E.}~\bibnamefont {Küçükbenli}}, \bibinfo {author} {\bibfnamefont {M.}~\bibnamefont {Lazzeri}}, \bibinfo {author} {\bibfnamefont {M.}~\bibnamefont {Marsili}},
  \bibinfo {author} {\bibfnamefont {N.}~\bibnamefont {Marzari}}, \bibinfo {author} {\bibfnamefont {F.}~\bibnamefont {Mauri}}, \bibinfo {author} {\bibfnamefont {N.~L.}\ \bibnamefont {Nguyen}}, \bibinfo {author} {\bibfnamefont {H.-V.}\ \bibnamefont {Nguyen}}, \bibinfo {author} {\bibfnamefont {A.}~\bibnamefont {{Otero-de-la-Roza}}}, \bibinfo {author} {\bibfnamefont {L.}~\bibnamefont {Paulatto}}, \bibinfo {author} {\bibfnamefont {S.}~\bibnamefont {Poncé}}, \bibinfo {author} {\bibfnamefont {D.}~\bibnamefont {Rocca}}, \bibinfo {author} {\bibfnamefont {R.}~\bibnamefont {Sabatini}}, \bibinfo {author} {\bibfnamefont {B.}~\bibnamefont {Santra}}, \bibinfo {author} {\bibfnamefont {M.}~\bibnamefont {Schlipf}}, \bibinfo {author} {\bibfnamefont {A.~P.}\ \bibnamefont {Seitsonen}}, \bibinfo {author} {\bibfnamefont {A.}~\bibnamefont {Smogunov}}, \bibinfo {author} {\bibfnamefont {I.}~\bibnamefont {Timrov}}, \bibinfo {author} {\bibfnamefont {T.}~\bibnamefont {Thonhauser}}, \bibinfo {author} {\bibfnamefont {P.}~\bibnamefont {Umari}}, \bibinfo {author} {\bibfnamefont {N.}~\bibnamefont {Vast}}, \bibinfo {author} {\bibfnamefont {X.}~\bibnamefont {Wu}},\ and\ \bibinfo {author} {\bibfnamefont {S.}~\bibnamefont {Baroni}},\ }\bibfield  {title} {\bibinfo {title} {Advanced capabilities for materials modelling with {{Quantum ESPRESSO}}},\ }\href {https://doi.org/10.1088/1361-648X/aa8f79} {\bibfield  {journal} {\bibinfo  {journal} {Journal of Physics: Condensed Matter}\ }\textbf {\bibinfo {volume} {29}},\ \bibinfo {pages} {465901} (\bibinfo {year} {2017})}\BibitemShut {NoStop}%
\bibitem [{\citenamefont {Perdew}\ \emph {et~al.}(1996)\citenamefont {Perdew}, \citenamefont {Burke},\ and\ \citenamefont {Ernzerhof}}]{perdew_1996}%
  \BibitemOpen
  \bibfield  {author} {\bibinfo {author} {\bibfnamefont {J.~P.}\ \bibnamefont {Perdew}}, \bibinfo {author} {\bibfnamefont {K.}~\bibnamefont {Burke}},\ and\ \bibinfo {author} {\bibfnamefont {M.}~\bibnamefont {Ernzerhof}},\ }\bibfield  {title} {\bibinfo {title} {Generalized {{Gradient Approximation Made Simple}}},\ }\href {https://doi.org/10.1103/PhysRevLett.77.3865} {\bibfield  {journal} {\bibinfo  {journal} {Physical Review Letters}\ }\textbf {\bibinfo {volume} {77}},\ \bibinfo {pages} {3865} (\bibinfo {year} {1996})}\BibitemShut {NoStop}%
\bibitem [{\citenamefont {Kresse}\ and\ \citenamefont {Furthmüller}(1996{\natexlab{a}})}]{kresse_1996}%
  \BibitemOpen
  \bibfield  {author} {\bibinfo {author} {\bibfnamefont {G.}~\bibnamefont {Kresse}}\ and\ \bibinfo {author} {\bibfnamefont {J.}~\bibnamefont {Furthmüller}},\ }\bibfield  {title} {\bibinfo {title} {Efficiency of ab-initio total energy calculations for metals and semiconductors using a plane-wave basis set},\ }\href {https://doi.org/10.1016/0927-0256(96)00008-0} {\bibfield  {journal} {\bibinfo  {journal} {Computational Materials Science}\ }\textbf {\bibinfo {volume} {6}},\ \bibinfo {pages} {15} (\bibinfo {year} {1996}{\natexlab{a}})}\BibitemShut {NoStop}%
\bibitem [{\citenamefont {Kresse}\ and\ \citenamefont {Furthmüller}(1996{\natexlab{b}})}]{kresse_1996a}%
  \BibitemOpen
  \bibfield  {author} {\bibinfo {author} {\bibfnamefont {G.}~\bibnamefont {Kresse}}\ and\ \bibinfo {author} {\bibfnamefont {J.}~\bibnamefont {Furthmüller}},\ }\bibfield  {title} {\bibinfo {title} {Efficient iterative schemes for ab initio total-energy calculations using a plane-wave basis set},\ }\href {https://doi.org/10.1103/PhysRevB.54.11169} {\bibfield  {journal} {\bibinfo  {journal} {Physical Review B}\ }\textbf {\bibinfo {volume} {54}},\ \bibinfo {pages} {11169} (\bibinfo {year} {1996}{\natexlab{b}})}\BibitemShut {NoStop}%
\bibitem [{\citenamefont {Iraola}\ \emph {et~al.}(2020)\citenamefont {Iraola}, \citenamefont {Mañes}, \citenamefont {Bradlyn}, \citenamefont {Neupert}, \citenamefont {Vergniory},\ and\ \citenamefont {Tsirkin}}]{iraola_2020}%
  \BibitemOpen
  \bibfield  {author} {\bibinfo {author} {\bibfnamefont {M.}~\bibnamefont {Iraola}}, \bibinfo {author} {\bibfnamefont {J.~L.}\ \bibnamefont {Mañes}}, \bibinfo {author} {\bibfnamefont {B.}~\bibnamefont {Bradlyn}}, \bibinfo {author} {\bibfnamefont {T.}~\bibnamefont {Neupert}}, \bibinfo {author} {\bibfnamefont {M.~G.}\ \bibnamefont {Vergniory}},\ and\ \bibinfo {author} {\bibfnamefont {S.~S.}\ \bibnamefont {Tsirkin}},\ }\bibfield  {title} {\bibinfo {title} {{{IrRep}}: Symmetry eigenvalues and irreducible representations of ab initio band structures},\ }\href@noop {} {\bibfield  {journal} {\bibinfo  {journal} {arXiv:2009.01764 [cond-mat, physics:physics]}\ } (\bibinfo {year} {2020})}\BibitemShut {NoStop}%
\bibitem [{\citenamefont {Aroyo}\ \emph {et~al.}(2006{\natexlab{a}})\citenamefont {Aroyo}, \citenamefont {{Perez-Mato}}, \citenamefont {Capillas}, \citenamefont {Kroumova}, \citenamefont {Ivantchev}, \citenamefont {Madariaga}, \citenamefont {Kirov},\ and\ \citenamefont {Wondratschek}}]{aroyo_2006a}%
  \BibitemOpen
  \bibfield  {author} {\bibinfo {author} {\bibfnamefont {M.~I.}\ \bibnamefont {Aroyo}}, \bibinfo {author} {\bibfnamefont {J.~M.}\ \bibnamefont {{Perez-Mato}}}, \bibinfo {author} {\bibfnamefont {C.}~\bibnamefont {Capillas}}, \bibinfo {author} {\bibfnamefont {E.}~\bibnamefont {Kroumova}}, \bibinfo {author} {\bibfnamefont {S.}~\bibnamefont {Ivantchev}}, \bibinfo {author} {\bibfnamefont {G.}~\bibnamefont {Madariaga}}, \bibinfo {author} {\bibfnamefont {A.}~\bibnamefont {Kirov}},\ and\ \bibinfo {author} {\bibfnamefont {H.}~\bibnamefont {Wondratschek}},\ }\bibfield  {title} {\bibinfo {title} {Bilbao {{Crystallographic Server}}: {{I}}. {{Databases}} and crystallographic computing programs},\ }\href {https://doi.org/10.1524/zkri.2006.221.1.15} {\bibfield  {journal} {\bibinfo  {journal} {Zeitschrift für Kristallographie - Crystalline Materials}\ }\textbf {\bibinfo {volume} {221}},\ \bibinfo {pages} {15} (\bibinfo {year} {2006}{\natexlab{a}})}\BibitemShut {NoStop}%
\bibitem [{\citenamefont {Aroyo}\ \emph {et~al.}(2006{\natexlab{b}})\citenamefont {Aroyo}, \citenamefont {Kirov}, \citenamefont {Capillas}, \citenamefont {{Perez-Mato}},\ and\ \citenamefont {Wondratschek}}]{aroyo_2006}%
  \BibitemOpen
  \bibfield  {author} {\bibinfo {author} {\bibfnamefont {M.~I.}\ \bibnamefont {Aroyo}}, \bibinfo {author} {\bibfnamefont {A.}~\bibnamefont {Kirov}}, \bibinfo {author} {\bibfnamefont {C.}~\bibnamefont {Capillas}}, \bibinfo {author} {\bibfnamefont {J.~M.}\ \bibnamefont {{Perez-Mato}}},\ and\ \bibinfo {author} {\bibfnamefont {H.}~\bibnamefont {Wondratschek}},\ }\bibfield  {title} {\bibinfo {title} {Bilbao {{Crystallographic Server}}. {{II}}. {{Representations}} of crystallographic point groups and space groups},\ }\href {https://doi.org/10.1107/S0108767305040286} {\bibfield  {journal} {\bibinfo  {journal} {Acta Crystallographica Section A: Foundations of Crystallography}\ }\textbf {\bibinfo {volume} {62}},\ \bibinfo {pages} {115} (\bibinfo {year} {2006}{\natexlab{b}})}\BibitemShut {NoStop}%
\bibitem [{\citenamefont {Aroyo}\ \emph {et~al.}(2011)\citenamefont {Aroyo}, \citenamefont {{Perez-Mato}}, \citenamefont {Orobengoa}, \citenamefont {Tasci}, \citenamefont {De~La~Flor},\ and\ \citenamefont {Kirov}}]{aroyo_2011}%
  \BibitemOpen
  \bibfield  {author} {\bibinfo {author} {\bibfnamefont {M.}~\bibnamefont {Aroyo}}, \bibinfo {author} {\bibfnamefont {J.}~\bibnamefont {{Perez-Mato}}}, \bibinfo {author} {\bibfnamefont {D.}~\bibnamefont {Orobengoa}}, \bibinfo {author} {\bibfnamefont {E.}~\bibnamefont {Tasci}}, \bibinfo {author} {\bibfnamefont {G.}~\bibnamefont {De~La~Flor}},\ and\ \bibinfo {author} {\bibfnamefont {A.}~\bibnamefont {Kirov}},\ }\bibfield  {title} {\bibinfo {title} {Crystallography online: {{Bilbao}} crystallographic server},\ }\href@noop {} {\bibfield  {journal} {\bibinfo  {journal} {Bulgarian Chemical Communications}\ }\textbf {\bibinfo {volume} {43}},\ \bibinfo {pages} {183} (\bibinfo {year} {2011})}\BibitemShut {NoStop}%
\end{thebibliography}%

\end{document}